\documentclass[11pt, reqno]{article}

\usepackage{jheppub}
\usepackage{amssymb}
\usepackage{amsmath}
\usepackage{amsthm}
\usepackage[usenames,dvipsnames]{xcolor}
\usepackage{epsfig}
\usepackage{dcolumn}
\usepackage{tikz}
\usetikzlibrary{shapes.geometric, arrows}
\usepackage{upgreek}
\usepackage{setspace}
\usepackage{enumitem}
\usepackage{array,multirow,bigdelim,arydshln}
\usepackage{appendix}
\usepackage{xparse}
\usepackage{stmaryrd}
\usepackage{graphicx}
\usepackage{float}
\usepackage{pifont}
\usepackage{xcolor}
\usepackage{mathtools}

\DeclarePairedDelimiter\floor{\lfloor}{\rfloor}

\usepackage[nottoc]{tocbibind}

\usepackage[utf8]{inputenc}
\hypersetup{
	colorlinks,
	urlcolor=Maroon,
	linkcolor=Maroon,
	citecolor=Maroon
}

\NewDocumentCommand{\binomial}{omm}
 {%
  \genfrac(){0pt}{}{#2}{#3}%
  \IfValueT{#1}{_{\!#1}}%
 }
\NewDocumentCommand{\eulerian}{omm}
 {%
  \genfrac<>{0pt}{}{#2}{#3}%
  \IfValueT{#1}{_{\!#1}}%
 }

\def \s {\sigma}

\usepackage{latexsym}
\usepackage{tikz}

\theoremstyle{plain}

\theoremstyle{definition}

\def\yz#1\yz {{\color{blue} YZ: #1 }}
\def\bu#1\bu {{\color{red} BU: #1 }}

\usepackage{tikz}

\title{Singular Solutions in Soft Limits}

\author[a]{Freddy Cachazo,}\emailAdd{fcachazo@pitp.ca}
\author[a,b]{Bruno Umbert,}\emailAdd{bgimenez@uwo.ca}
\author[c,d,a]{and Yong Zhang}\emailAdd{yongzhang@itp.ac.cn}

\affiliation[a]{Perimeter Institute for Theoretical Physics, Waterloo, ON N2L 2Y5, Canada}
\affiliation[b]{Department of Applied Mathematics, Western University, London, ON N6A 5B7, Canada}
\affiliation[c]{CAS Key Laboratory of Theoretical Physics, Institute of Theoretical Physics, Chinese Academy of Sciences, Beijing 100190, China}
\affiliation[d]{School of Physical Sciences, University of Chinese Academy of Sciences, No.19A Yuquan Road, Beijing 100049, China}

\abstract{A generalization of the scattering equations on $X(2,n)$, the configuration space of $n$ points on $\mathbb{CP}^1$, to higher dimensional projective spaces was recently introduced by Early, Guevara, Mizera, and one of the authors. One of the new features in $X(k,n)$ with $k>2$ is the presence of both regular and singular solutions in a soft limit. In this work we study soft limits in $X(3,7)$, $X(4,7)$, $X(3,8)$ and $X(5,8)$, find all singular solutions, and show their geometrical configurations. More explicitly, for $X(3,7)$ and $X(4,7)$ we find $180$ and $120$ singular solutions which when added to the known number of regular solutions both give rise to $1\, 272$ solutions as it is expected since $X(3,7)\sim X(4,7)$. Likewise, for $X(3,8)$ and $X(5,8)$ we find $59\, 640$ and $58\, 800$ singular solutions which when added to the regular solutions both give rise to $188\, 112$ solutions. We also propose a classification of all configurations that can support singular solutions for general $X(k,n)$ and comment on their contribution to soft expansions of generalized biadjoint amplitudes.
}

\begin{document}
\maketitle
\addtocontents{toc}{\protect\setcounter{tocdepth}{1}}
\def \tr {\nonumber\\}
\def \la  {\langle}
\def \ra {\rangle}
\def\hset{\texttt{h}}
\def\gset{\texttt{g}}
\def\sset{\texttt{s}}
\def \be {\begin{equation}}
\def \ee {\end{equation}}
\def \ba {\begin{eqnarray}}
\def \ea {\end{eqnarray}}
\def \k {\kappa}
\def \h {\hbar}
\def \r {\rho}
\def \l {\lambda}
\def \be {\begin{equation}}
\def \en {\end{equation}}
\def \bes {\begin{eqnarray}}
\def \ens {\end{eqnarray}}
\def \red {\color{Maroon}}
\def \pt {{\rm PT}}
\def \s {\sigma} 
\def \ls {{\rm LS}}
\def \ma {\Upsilon}
\def \s {\textsf{s}}
\def \t {\textsf{t}}
\def \R {\textsf{R}}
\def \W {\textsf{W}}
\def \U {\textsf{U}}
\def \e {\textsf{e}}

\numberwithin{equation}{section}

\section{Introduction}

Recently Early, Guevara, Mizera, and one of the authors introduced and studied a natural generalization of the scattering equations, which connect the space of Mandelstam invariants to that of points on ${\mathbb{CP}^1}$ \cite{Fairlie:1972zz,Fairlie:2008dg,Cachazo:2013gna,Cachazo:2013hca}, to higher dimensional projective spaces $\mathbb{CP}^{k-1}$ \cite{CEGM}. The equations are obtained by computing the critical points of a potential function
\be\label{gen}
{\cal S}_k \equiv \sum_{1\leq a_1<a_2 \cdots <a_k\leq n} s_{a_1a_2\cdots a_k}\log\,( a_1,a_2,\ldots ,a_k).
\ee
Here $s_{a_1a_2\cdots a_k}$ are a generalization of Mandelstam invariants while $( a_1,a_2,\cdots ,a_k)$ can be thought of as Pl\"ucker coordinates on $G(k,n)$. The configuration space of $n$ points on $\mathbb{CP}^{k-1}$ is obtained by modding out by a torus action $\mathbb{C}^*$ on each of the points, i.e., $X(k,n):= G(k,n)/(\mathbb{C}^*)^n$ \cite{sekiguchi1997w}.

The kinematic invariants are completely symmetric tensors satisfying
\be
s_{aabc\cdots} = 0, \qquad \sum_{a_2,a_3,\ldots ,a_k}s_{a_1a_2\cdots a_k} = 0 \qquad \forall\, a_1.
\ee
These are the analogs of masslessness and momentum conservation conditions. These conditions guarantee that the potential function is invariant under the torus action and therefore one can choose inhomogeneous coordinates for points on $\mathbb{CP}^{k-1}$. For example, when $k=3$ one can use $(x_i,y_i)$ while the Pl\"ucker coordinates are then replaced by
\be
|abc| := {\rm det}\left(
                    \begin{array}{ccc}
                      1 & 1 & 1 \\
                      x_a & x_b & x_c \\
                      y_a & y_b & y_c \\
                    \end{array}
                  \right).
\ee

Having a higher $k$ version of the scattering equations, the most natural question is to determine the number of solutions, i.e. the number of critical points of the potential ${\cal S}_k$. The standard scattering equations, i.e. $k=2$, possess $(n-3)!$ solutions and the original proof given in \cite{Cachazo:2013iea} uses that a soft particle decouples from the rest and proceeds by induction. The argument relies on the fact that as the soft limit is approached, all solutions stay away from boundaries of $X(2,n)$, i.e. the $n$ points are in a generic configuration. These solutions are known as {\it regular} solutions. The terminology comes from the study of factorization limits, i.e. when a physical kinematic invariant vanishes. In such a limit, some solutions give rise to configurations where the Riemann sphere degenerates into two spheres joined by a single, emergent puncture. Such solutions are called {\it singular} solutions.

In \cite{CEGM} it was found that when $k\geq3$ regular solutions in a soft limit cannot possibly account for all solutions. This was deduced by computing the regular solutions for $X(3,7)\to X(3,6)$ and $X(4,7)\to X(4,6)$. The numbers were shown to be $1\,092$ and $1\,152$ respectively. Since $X(3,7)$ and $X(4,7)$ are isomorphic, they must possess the same number of total solutions. Motivated by this, Rojas and one of the authors designed a technique for determining the number of missing solutions for $X(4,7)$ as the rank of a matrix built out of generalized biadjoint amplitudes thus finding $120$ \cite{Cachazo:2019apa}. This implies that the total number of solutions is exactly $1\, 272$ and that the number of singular solutions for $X(3,7)$ and $X(4,7)$ must be $180$ and $120$ respectively.

For $X(3,8)\to X(3,7)$ and $X(5,8)\to X(5,7)$ one can also compute the number of regular solutions and find them to be $128\,472$ and $129\,312$ respectively. Once again, since $X(3,8)$ is isomorphic to $X(5,8)$ there must be singular solutions. At this point there is no technique for computing the total number of solutions from the scattering equations or generalized biadjoint scalar amplitudes. However, Lam has suggested that the total number of solutions can be related to the number of representations of uniform matroids over finite fields\footnote{Private communication.}. Using this one can reproduce the correct number for $X(2,n)$, $X(3,6)$, and $X(3,7)$. Moreover, it also predicts $188\,112$ solutions for $X(3,8)$.

In this work we study the soft limits of scattering equations on $X(3,7)$, $X(4,7)$, $X(3,8)$, and $X(5,8)$ and find all singular solutions. In each case, singular solutions correspond to configurations where the soft particle develops some linear dependence with subsets of the hard particles while every minor containing only hard particles remains finite. Such linear dependencies prevent the decoupling of the soft particle from the rest. The simplest example corresponds to $X(3,7)$ when particle $7$ is taken to be soft and a configuration where $|147|$, $|257|$ and $|367|$ vanish. This means that the terms containing $s_{147}$, $s_{257}$ and $s_{367}$ cannot be dropped in the scattering equations for the hard particles as it is usually the case for regular solutions.

We find that in every case it is possible to define a new set of scattering equations in the strict soft limit. This is a completely novel phenomenon. The strict soft limit scattering equations can be solved or its solutions counted using some of the same techniques developed for the original scattering equations. In fact, using a soft-limit approach one finds again regular and singular solutions.

Based on these examples we propose a general classification of all configurations that can support singular solutions in $X(k,n)$ for general $k$ and $n$. For example, when $k=3$ there are $\floor*{\frac{n-1}{2}}-2$ distinct topologies corresponding to $3$, ..., $\floor*{\frac{n-1}{2}}$ lines intersecting at the soft particle position. For higher $k$, there are configurations that are inherited from lower $k$ values as well as new ones corresponding to at least $k$ $(k-2)$-planes intersecting at the soft particle location. The general structure hints at a recursive structure for $X(k,n)$ similar to that found for $X(2,n)$.

Very recently, an elegant structure of soft theorems was unearthed by Garcia and Guevara in generalized biadjoint amplitudes \cite{GG}. One of the surprising results is the fact that standard $k=2$ biadjoint amplitudes serve as soft factors for $k>2$ amplitudes. They computed the leading order behavior of amplitudes in the soft limit, i.e., as $\tau\to 0$ with $s_{abn}=\tau {\hat s}_{abn}$, assuming a decoupling of the soft particle from the scattering equations governing the hard particles. We find that in all examples we studied their assumption is indeed correct as singular solutions can at most contribute to subleading terms in the soft limit expansion.

This paper is organized as follows: In section \ref{sec2} we review the standard argument for $k=2$ adding an explanation for why no singular solutions are found. In section \ref{sec3} we review what it is known regarding regular solutions, in particular, how this led  to the prediction of the existence of singular solutions. In section \ref{sec4} we find all singular solutions in the soft limits $X(3,7)\to X(3,6)$ and $X(4,7)\to X(4,6)$. In section \ref{sec5} we find all singular solutions in the soft limits $X(3,8)\to X(3,7)$ and $X(5,8)\to X(5,7)$. In the latter we find for the first time topologically distinct configurations leading to singular solutions. In section \ref{sec6}, we make our proposal for all configurations that can support singular solutions and explain the evidence supporting it. We end in section \ref{sec7} with discussions regarding the contribution of singular solutions to the soft expansion of generalized biadjoint scalar amplitudes. In appendix \ref{appendix1} we show how the counting of the number of singular solutions works from the bounded chambers method in some particular cases for $k=3$, and in appendix \ref{appendix2} we comment on the geometrical interpretation of some of the singular configurations in $X(5,8)$.

\section{Soft Limits in $X(2,n)$\label{sec2}}

Scattering equations on $X(2,n)$ have provided a direct connection between locality and unitarity constraints in tree-level scattering amplitudes and properties of the moduli space of punctured Riemann spheres. The way this happens is somewhat surprising. The scattering equations for $n$ particles possess ${\cal N}_n = (n-3)!$ solutions and when a factorization channel, in which particles separate into two sets $L$, $R$, containing $n_L>1$ and $n_R>1$ particles, is approached, ${\cal N}_n^{\rm  singular}:= (n_L-2)!\times (n_R-2)!$ solutions become singular. More explicitly, all punctures in $L$ (or R) approach each other\footnote{This is in some ${\rm SL}(2,\mathbb{C})$ gauge choice.}. However, cross ratios involving only particles on $L$ (or $R$) remain finite and lead to the blow up picture where two Riemann spheres are joined by a new puncture with one containing particles in $L$ and the other particles in $R$.

The singular solutions are the most relevant to ensure the correct physical behavior of scattering amplitudes in the Cachazo-He-Yuan (CHY) formulation as they produce the kinematic pole while the remaining ${\cal N}_n^{\rm regular}:=(n-3)!-{\cal N}_n^{\rm  singular}$ are regular. This means that the CHY formula remains finite on them. This is precisely the opposite to what happens in a soft limit. Indeed, in $X(2,n)$ one finds only regular solutions and they are the ones responsible for the leading order behavior of amplitudes in the limit and control the corresponding soft theorems \cite{Weinberg:1965nx,Cachazo:2014fwa,Schwab:2014xua,Afkhami-Jeddi:2014fia}. In this section we review the soft limit analysis as preparation for $X(k,n)$ with $k>2$.

\subsection{Regular Solutions}

Let us write the scattering equations in a form that manifestly exhibits the dependence on particle $n$:
\be\label{softi}
E_a:=\sum_{b=1}^{n-1}\frac{s_{ab}}{x_{ab}} + \frac{s_{an}}{x_{an}}\quad {\rm with}\quad 1\leq a \leq n-1 \quad {\rm and} \quad E_n:= \sum_{b=1}^{n-1}\frac{s_{nb}}{x_{nb}}
\ee
with $x_{ab}=x_a-x_b$ and the equations are obtained by requiring $E_a=0$ for all $a$.

The soft limit in particle $n$ is defined by taking $s_{an}=\tau {\hat s}_{an}$ with $\tau\to 0$. Regular solutions are defined as those where none of the punctures approach another. More explicitly, $x_{ab}\neq 0$ for all values of $a$ and $b$. Under this assumption it is easy to see from \eqref{softi} that all $n$ dependence can be dropped from the first $n-1$ equations. This set of equations precisely corresponds to that of a system of $n-1$ particles and therefore can be solved to find ${\cal N}_{n-1}$ solutions. In other words, in the soft limit, the $n^{\rm th}$ particle decouples from the equations that control the rest. However, the possible values of $x_n$ are not arbitrary since $\tau$ drops from the last equation in \eqref{softi} to give
\be\label{wimo}
\sum_{b=1}^{n-1}\frac{{\hat s}_{nb}}{x_{n}-x_b^{I}}=0
\ee
where $x_b^I$ is any one of the ${\cal N}_{n-1}$ solutions for the hard particles. At first sight this equation leads to a polynomial in $x_n$ of degree $n-2$ but the coefficient of $x_n^{n-2}$ vanishes due to momentum conservation and hence it leads to $n-3$ solutions for $x_n$. Since this is true for each $x_b^I$ one finds ${\cal N}_n^{\rm regular} = (n-3){\cal N}_{n-1}$.

Under the assumption that ${\cal N}_n^{\rm  singular} = 0$ one finds the recursion relation ${\cal N}_n = (n-3){\cal N}_{n-1}$ with ${\cal N}_4=1$ and whose solution is ${\cal N}_n = (n-3)!$. Now we turn to proving that ${\cal N}_n^{\rm  singular} = 0$.

\subsection{Absence of Singular Solutions}

A singular solution is one which does not obey the condition for decoupling the soft particle from the equations determining the rest. This can only happen when $x_{in}= \tau {\hat x}_{in}$, i.e. vanishes in the soft limit for some values of $i$. Let us denote the set of such particles ${\cal D}$. Clearly ${\cal D}$ must contain more than one element for if $|{\cal D}|=1$ then the last equation in \eqref{softi} becomes $E_n = {\hat s}_{in}/{\hat x}_{in} = 0$ which has no solutions.

Let us assume that $|{\cal D}|\geq 2$ and parameterize $x_{i} = x_n + \tau u_i$ for $i\in {\cal D}$. Here we follow an argument originally presented in \cite{Cachazo:2013gna} for factorization limits but perfectly applicable to the situation at hand. It is simple to show that for any $a\notin {\cal D}$
\be
x_{an}E_a = \sum_{b \notin {\cal D}}\frac{x_{an}}{x_{ab}}s_{ab}+\sum_{b\in {\cal D}}\left(1+\tau\frac{u_{b}}{x_{ab}}\right)s_{ab}.
\ee
Of course, this must be zero when the scattering equations are imposed. Adding all these equations one finds
\be\label{contra}
\sum_{a\notin {\cal D}}x_{an}E_a=0\quad \Rightarrow\quad \left(\sum_{a\notin {\cal D}}k_{a}\right)^2 = {\cal O}(\tau).
\ee
However, an implicit assumption in a soft limit is that the kinematics of the system of $n-1$ particles is generic and therefore no kinematic invariant involving only hard particles is allowed to vanish. This means that \eqref{contra} is a contradiction and therefore singular solutions do not exist in the soft limit $X(2,n)\to X(2,n-1)$.

\section{Regular Solutions in $X(k,n)\to X(k,n-1)$}\label{sec3}

In this section we review the known results for the counting of regular solutions in the soft limits $X(k,n)\to X(k,n-1)$. As discussed in the previous section, regular solutions are defined as those for which the soft particle decouples from the equations determining the configuration of the others. This means that we can assume that the system $X(k,n-1)$ has been solved and ${\cal N}_{n-1}^{(k)}$ solutions have been found. The task at hand is then to determine the number of solutions for the position of particle $n$ from the equations
\be\label{resul}
\triangledown_n {\cal S}_{k} = 0.
\ee
Here the gradient is taken only with respect to the coordinates of particle $n$ since all other particle positions are assumed to have been found. Let us denote the number of solutions to \eqref{resul} as ${\rm Soft}_{k,n}$. The notation is motivated by soft theorems. This means that the number of regular solutions is ${\cal N}^{(k):{\rm regular}}_{n} = {\rm Soft}_{k,n}\times {\cal N}_{n-1}^{(k)}$.

In the soft limit $X(2,n)\to X(2,n-1)$ we have seen that \eqref{resul} is a single equation with ${\rm Soft}_{2,n}=n-3$ solutions and therefore ${\cal N}^{(2):{\rm regular}}_{n} = (n-3)\times {\cal N}^{(2)}_{n-1}$. Of course, we have seen that ${\cal N}^{(2):{\rm regular}}_{n}$ is also equal to the total number of solutions ${\cal N}^{(2)}_{n}$.

The only other case that is known for all $n$ is the soft limit $X(3,n)\to X(3,n-1)$. In \cite{CEGM} it was found that
\be
{\rm Soft}_{3,n}=\frac{1}{8}(n-4)(n^3-6n^2+11n-14).
\ee
The first few values are ${\rm Soft}_{3,5}=2,{\rm Soft}_{3,6}=13,{\rm Soft}_{3,7}=42$, and ${\rm Soft}_{3,8}=101$. By explicit computations it was found in \cite{CEGM} that ${\cal N}^{(3)}_{5}=2$ and ${\cal N}^{(3)}_{6}=2\times 13=26$. This means that there are no singular solutions for $n\leq 6$. Therefore the number of regular solutions for $n=7$ is ${\cal N}^{(3):{\rm regular}}_{7}=42\times 26 = 1\,092$. In \cite{Cachazo:2019apa}, it was proven that the total number of solutions for $n=7$ is ${\cal N}^{(3)}_{7}=1\,272$ and with this the number of regular solutions in the soft limit $X(3,8)\to X(3,7)$ is ${\cal N}^{(3):{\rm regular}}_{8}=101\times 1\,272 =128\,472$. In section \ref{sec5} we show that the total number of solutions for $X(3,8)$ is ${\cal N}^{(3)}_{8}=188\,112$. Therefore the number of regular solutions for $n=9$ is ${\cal N}^{(3):{\rm regular}}_{9}=205\times 188\,112 = 38\,562\,960$. Since the total number of solutions for $X(3,9)$ is not presently known we cannot determine ${\cal N}^{(3):{\rm regular}}_{n}$ for $n\geq 10$.

In \cite{CEGM}, the number of regular solutions was identified with the number of bounded chambers by real hyperplanes when the kinematics was chosen in a special region known as the positive region (reviewed in section \ref{sec411}). This identification is also based on the assumption that all solutions are real in the positive region. Using this approach ${\rm Soft}_{4,6}=6$, ${\rm Soft}_{4,7}=192$ and ${\rm Soft}_{4,8}=1\,858$ were computed. Here we have pushed the computation of bounded chambers up to $n=16$ leading to the following proposal
\be
{\rm Soft}_{4,n}= \frac{1}{1296} (n-5 ) (n^8- 13 n^7 - 5 n^6 +1019 n^5- 7934 n^4  + 29198 n^3- 57510 n^2+ 57276 n -20736 ).
\ee
These results imply that
\be
{\cal N}^{(4):{\rm regular}}_{7}=6\times 192=1\, 152,\quad {\cal N}^{(4):{\rm regular}}_{8}= 1\,272\times 1\,858=2\, 363\, 376.
\ee
For $k=5$ much less is known: ${\rm Soft}_{5,7}=24$,  ${\rm Soft}_{5,8}=5\,388$ and  ${\rm Soft}_{5,9}=204\,117$. This leads to
\be
{\cal N}^{(5):{\rm regular}}_{8}=24\times 5\,388=129\, 312,\quad {\cal N}^{(5):{\rm regular}}_{9} =204\,117\times 188\,112.
\ee
The last result uses that the total number of solutions of the scattering equations on $X(5,8)$ is ${\cal N}^{(5)}_{8} = {\cal N}^{(3)}_{8} = 188\,112$.

\section{Singular Solutions in $X(3,7)\to X(3,6)$ and $X(4,7)\to X(4,6)$} \label{sec4}

We have already seen that there cannot be singular solutions for $k=2$. For higher $k$, however, it is possible to keep all minors without the soft particle finite while sending some of the minors involving the soft particle to zero. This makes singular solutions possible for $k>2$. In this section we study the first examples where singular solutions appear, which correspond to $X(3,7)\to X(3,6)$ and $X(4,7)\to X(4,6)$. This analysis also explains why there are not singular solutions for $X(3,6)\to X(3,5)$ explaining the agreement of the regular soft counting of solutions with the total number of solutions found in \cite{CEGM}.

\subsection{Singular Solutions in $X(3,7)\to X(3,6)$} \label{x37}

The first explicit example where we have singular solutions is in $X(3,7)$. In order to obtain the singular solutions, we study the soft limit for, e.g., particle $n=7$, {\it i.e.}  $s_{ab7}\to\tau \hat{s}_{ab7}$ (with $\tau\to 0$).  The singular solutions arise from configurations where three lines\footnote{In this work we use the word ``line" to refer to a complex line, i.e., $\mathbb{CP}^1$, or to a real line. The meaning should be clear from the context.} in $\mathbb{CP}^2$ (or $\mathbb{RP}^2$ if all solutions are real), each defined by two hard particles, meet at the soft particle.

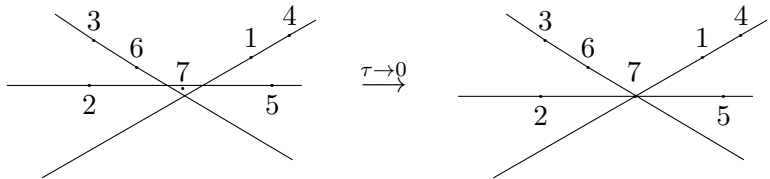
\begin{figure}[!htb]
\centering
\begin{tikzpicture} [xscale=1,yscale=1]
\begin{scope}
\draw (4.20,-1.18)--(4.72,-1.53) node {.} node[above] {3}
--(5.29,-1.89)node {.}  node[above] {6}
--(7.36,-3.12)

(3.56,-2.13)--(4.66,-2.13)node{.}node[below]{2}--(7.09,-2.13)node{.}node[below]{5}--
(7.48,-2.13)

(4.03,-3.36)--(6.81,-1.76)node{.}node[above]{1}--(7.32,-1.46)node{.}node[above]{4}--(7.67,-1.26)
;
\node at (5.90,-2.17) {.} ;

\node at (5.90,-1.95) {7};

\node at (8.57,-2.) {$\overset{\tau\to0}{\longrightarrow}$};
\end{scope}

\begin{scope}[yshift=0cm,xshift=6cm]
\draw (4.20,-1.18)--(4.72,-1.53) node {.} node[above] {3}
--(5.29,-1.89)node {.}  node[above] {6}
--(7.36,-3.12)

(3.56,-2.275)--(4.66,-2.275)node{.}node[below]{2}--(7.09,-2.275)node{.}node[below]{5}--
(7.48,-2.275)

(4.03,-3.36)--(6.81,-1.76)node{.}node[above]{1}--(7.32,-1.46)node{.}node[above]{4}--(7.67,-1.26)
;
\node at (5.92,-2.275) {.} ;

\node at (5.90,-1.95) {7};

\end{scope}

\end{tikzpicture}

\caption{Configuration of singular solutions in $X(3,7)$. \textit{Left}: Near the soft limit three lines $\overline{14}$, $\overline{25}$ and $\overline{36}$ almost cross the soft particle. \textit{Right}: In the strict soft limit the three lines meet at the soft particle. \label{x37t}}
\end{figure}

One such configuration is where lines $\overline{14}$, $\overline{25}$ and $\overline{36}$ meet at the particle 7  as shown in figure \ref{x37t}. This implies that all three determinants $|147|$, $|257|$ and $|367|$ vanish. There exist ${6 \choose 2}{4 \choose 2}{4 \choose 2}/3!=15$ different such configurations.

For each configuration, it is possible to choose coordinates to find equations governing the system at $\tau=0$. The new scattering equations have 12 solutions. Therefore there are ${\cal N}_7^{(3):{\rm singular}}=12\times 15=180$ singular solutions.

The way  to get the solutions is the following. Take the configuration where $|147|$, $|257|$ and $|367|$ vanish as an example.  A convenient choice of gauge fixing in projective space is
\be\label{gauge}
\left(
                    \begin{array}{ccccccc}
                      1 & 0 & 0 &1    &1   & 1 & 1 \\
                      x_{1} & x_{2} & x_{3}   & x_{4} &x_{5}   &x_{6}  &x_{7}  \\
                      y_{1} & y_{2} & y_{3}  & y_{4}    &y_{5}   &y_{6}  &y_{7}   \\
                    \end{array}
                  \right)\overset{\text{gauge~fixing}}{\to}
\left(
                    \begin{array}{ccccccc}
                      1 & 0 & 0 &1    &1   & 1 & 1 \\
                      0 & 1 & 0   & 1 &x_{5}   &x_{6}  &x_{7}  \\
                      0 & 0 & 1  & 1    &y_{5}   &y_{6}  &y_{7}   \\
                    \end{array}
                  \right).
\ee
Under the parametrization $s_{ab7}\to\tau \hat{s}_{ab7}$, terms containing $s_{147}$, $s_{257}$ and $s_{367}$ cannot be dropped in the equations for the hard particles
\be\label{neweqA}
  \frac{\partial {\tilde{\cal S}}_{3}}{\partial x_{a} } + \sum_{b\neq a,7} \frac{\tau \hat{s}_{ab7}}{|ab7|} \frac{\partial |ab7|}{\partial x_{a}} =0, \quad   \frac{\partial {\tilde{\cal S}}_{3}}{\partial y_{a} } + \sum_{b\neq a,7} \frac{\tau \hat{s}_{ab7}}{|ab7|} \frac{\partial |ab7|}{\partial y_{a}}=0,  \quad \text{for ~} a=1,\ldots 6
\ee
where ${\tilde{\cal S}_{3}}$ is the potential of hard particles, $
{\tilde{\cal S}_{3}}\equiv \sum_{1\leq a<b<c\leq 6}s_{abc}\log (a, b , c)
$. They also dominate in the two scattering equations for the soft particle
\begin{align} 
  \frac{\partial {\cal S}_{3}}{\partial x_{7} }= 
 \frac{\tau {\hat s}_{147}}{|147|} \frac{\partial |147|}{\partial x_{7}} + \frac{\tau {\hat s}_{257}}{|257|} \frac{\partial |257|}{\partial x_{7}} + \frac{ \tau {\hat s}_{367}}{|367|} \frac{\partial |367|}{\partial x_{7}}+ {\cal O}(\tau)=0 \,,
 \nonumber
\\ 
  \frac{\partial {\cal S}_{3}}{\partial y_{7} }= 
 \frac{ \tau {\hat s}_{147}}{|147|} \frac{\partial |147|}{\partial y_{7}} + \frac{\tau {\hat s}_{257}}{|257|} \frac{\partial |257|}{\partial y_{7}} + \frac{ \tau {\hat s}_{367}}{|367|} \frac{\partial |367|}{\partial y_{7}}+ {\cal O}(\tau)=0 \,.  \label{soft}
  \end{align}
The subleading terms ${\cal O}(\tau)$ in the above equations \eqref{neweqA} and \eqref{soft} can be omitted in the soft limit\footnote{This is because the terms shown explicitly in \eqref{soft} are of order ${\cal O}(\tau^0)$ since the minors in the denominators vanish as ${\cal O}(\tau)$ thus canceling the explicit factor of $\tau$ in the numerators.}. In contrast to regular solutions, where we solve the equations for hard particles first, here the equations for the soft particle \eqref{soft} are simpler and we solve them first. Note that there are three dominating terms in each of the equations \eqref{soft}. Algebraically, one can check that there would be no solutions for $x_7$ and $y_7$ if there were only two dominating terms in each of the equations \eqref{soft}. In fact, this is the reason why there are no singular solutions for $X(3,6)\to X(3,5)$. The fact that at least three terms are needed has a more intuitive geometric explanation which we give in the next subsection. 

We then parametrize each determinant as $|147|=\tau u$, $|257|=\tau v$ and $|367|=\tau p$, that is
\be\label{para}
  x_{6} = y_{5} - \tau ( u  + v + p),\quad
x_{7} = y_{5} -  \tau ( u + v) ,  \quad y_{7} = y_{5} -\tau v.
\ee
In the soft limit,
the new set of scattering equations, with variables $x_{5}$, $y_{5}$, $y_{6}$, $u$, $v$ and $p$ is 
\be\label{neweq}
\left. \lim_{\tau\to0} \, \frac{\partial {\cal S}_{3}}{\partial x_{i}}\right| _{\eqref{para}}=0, \left. \quad  \lim_{\tau\to0} \,  \frac{\partial {\cal S}_{3}}{\partial y_{i}}\right| _{\eqref{para}}=0,  \quad \text{for ~} i=1,\ldots 7.
\ee
Among the 14 equations \eqref{neweq}, only 6 of them are independent. Furthermore, we can separately solve for $u$, $v$ and $p$ from \eqref{neweq} and obtain equations involving only hard particles
\be\label{neweq37}
\left. \Big(\frac{\partial \tilde{\cal S}_{3}}{\partial y_{5}}+ \frac{\partial \tilde{\cal S}_{3}}{\partial x_{6}}\Big)\right|_{x_{6}\to y_{5}}=0, \left. \quad  \frac{\partial \tilde{\cal S}_{3}}{\partial x_{5}} \right|_{x_{6}\to y_{5}}=0, \left. \quad  \frac{\partial \tilde{\cal S}_{3}}{\partial y_{6}} \right|_{x_{6}\to y_{5}}=0\,.
\ee
Solving these equations we find that, compared to the original scattering equations for 6 particles, which have 26 solutions, now the requirement that lines $\overline{14}$, $\overline{25}$ and $\overline{36}$ pass through a common point reduces the number of solutions to $12$.  

\subsubsection{Singular Solutions on Positive Kinematics}
\label{sec411}
We have seen the kind of configurations that produce singular solutions in $X(3,7)\to X(3,6)$. However, a purely algebraic approach sheds little light on why such configurations can produce singular solutions while others cannot. Moreover, unless a more geometric understanding is reached, it seems hopeless to uncover the general structure for all soft limits $X(k,n)\to X(k,n-1)$.

In this subsection, we make use of kinematic data in what is known as the positive region ${\cal K}_{3,n}^+$ to study and visualize the solutions (for more the details on ${\cal K}_{3,n}^+$ see \cite{Cachazo:2016ror,CEGM}). The main advantage is that one can develop intuition on why there are singular solutions through explicit geometric pictures. 

Let us briefly review the construction of  
kinematic data in the positive region ${\cal K}_{k,n}^+$ for general $k$.   
 We start by selecting $k+1$ particles $A_{1}$, $A_{2}$, $\cdots$, $A_{k+1}$ to be fixed by the action of $\mathrm{SL}(k, \mathbb{C})$. This time $k-1$ particles, say $A_{2}$, $A_{3}$, $\cdots$, $A_{k}$, can be sent to infinity in $k-1$ different directions by setting their homogeneous coordinates to $(0, 1, 0,\cdots,0)$, $(0,0,1,\cdots,0)$ , $\cdots$, $(0,0,\cdots, 0, 1)$, respectively.
The other two are chosen to be, in inhomogeneous coordinates, at the origin and at $(1, 1,\cdots,1)$ on the plane $(x_{1}, x_{2},\cdots,x_{k-1}) \in {\mathbb R}^{k-1}$.

Since interactions in the potential function are controlled by the determinants $|a_{1}a_{2}\cdots a_{k}|$, a given particle is not directly sensitive to the location of any other particle but only sensitive to the $(k-2)$-planes defined by any other $k-1$ particles. In order to find the analog of the positive region, let us again consider the potential function
\begin{align}\label{potential4}
\mathcal{S}_{k} =  \!\!\!\!\!\!\!\! \sum_{\substack{1\leq a_{1}<a_{2}<\cdots<a_{k} \leq n  \\ |\{a_{1},a_{2},\cdots,a_{k}\}\cap \{A_{1}, A_{2}, \cdots,A_{k+1}\}|\leq k-1
\\
|\{a_{1},a_{2},\cdots,a_{k}\}\cap \{A_{2}, A_{3}, \cdots,A_{k}\}|\leq k-
2}}  \!\!\!\!\!\!\!\! s_{a_{1},a_{2},\cdots,a_{k}} \log |  a_{1},a_{2},\cdots,a_{k} |\,.
\end{align}
Therefore, this positive region ${\cal K}_{k,n}^{+}$ is defined by requiring all invariants that explicitly appear in \eqref{potential4} to be positive. This is possible because the set of all such invariants form a basis of the kinematic space. Since critical points of the potential correspond to equilibrium points, they can only lie inside the bounded chambers of this space, assuming they are all real.

Let us define the subregion of ${\cal K}_{k,n}^{+}$ where all solutions to the scattering equuations are real by ${\cal K}_{k,n}^{+,\mathbb{R}}$.

When $k=2$, it is known that ${\cal K}_{2,n}^{+,\mathbb{R}}= {\cal K}_{2,n}^{+}$. Moreover, since ${\cal K}_{2,n}^{+}$ contains all soft limits, it is possible to smoothly go from one to another without ever leaving  ${\cal K}_{2,n}^{+,\mathbb{R}}$. In \cite{CEGM}, it was argued that for $k=3$ it turns out that ${\cal K}_{3,n}^{+,\mathbb{R}}\subset {\cal K}_{3,n}^{+}$ is disconnected. In fact, each soft limit seems to live in its own region. For our present problem of $X(3,7)$, it is enough to know that sufficiently near the soft limit of particle $7$ all solutions are real. 

Singular solutions are called singular because they make some minors $|ab7|$ containing particle 7 vanish. Geometrically, this means that lines  $\overline{ab}$ in  ${\mathbb {RP}}^{2}$ space  will dominate. The remaining lines can be omitted for the soft particle at first. Therefore, in order to bound particle $7$ in  ${\mathbb {RP}}^{2}$ space, we need at least $3$ such dominating lines. That is, we need at least three vanishing minors involving particle $7$ while keeping the other minors still finite. For $n=7$, this can be achieved for example by letting $|147|$, $|257|$ and $|367|$ vanish. 
 There are 15 such kind of configurations. In  appendix \ref{appendix12},  we further find out that there are 12 bounded chambers to bound the soft particle 7,  which means there are 12 solutions for each of the configurations.

\subsection{Singular Solutions in $X(4,7)\to X(4,6)$} \label{x47777}

\begin{figure}[!htb]
\centering
\begin{tikzpicture} [xscale=1,yscale=1]
\begin{scope}
\draw (3.74,-1.69)-- (5.25,-1.74);
\draw[dashed](5.25,-1.74)--(6.26,-1.77);
\draw(6.26,-1.77)--
(7.37,-1.80) node {.} node[below] {4}
--(7.64,-1.81);

\draw[blue] (3.65,-2.94)--(6.12,-1.09) node {\color{black}.} node[above left=-3pt] {\color{black}{6}}
--(7.42,-0.13);

\draw[red] (3.94,-2.89)--(7.51,-1.16) node {\color{black}.} node[above] {\color{black}{3}}
--(7.80,-1.01);

\draw[purple] (4.30,-0.14)--(4.77,-0.52) node {\color{black}.} node[above] {\color{black}{2}}
--(7.36,-2.68);

\draw[gray] (5.78,-0.38)--(5.78,-0.62) node {\color{black}.} node[left] {\color{black}{1}}
--(5.79,-3.58);

\draw[green] (3.46,-0.78)--(6.67,-2.50) node {\color{black}.} node[below] {\color{black}{5}}
--(7.54,-2.95);

\node at (5.65,-1.84) {.} ;

\node at (5.37,-2.00) {7};

\node at (8.57,-1.74) {$\overset{\tau\to0}{\longrightarrow}$};
\end{scope}

\begin{scope}[yshift=0cm,xshift=6cm]
\draw (3.74,-1.69)--
(7.37,-1.80) node {.} node[below] {4}
--(7.64,-1.81);

\draw[blue] (3.65,-2.94)--(6.12,-1.09) node {\color{black}.} node[above left=-3pt] {\color{black}{6}}
--(7.42,-0.13);

\draw[red] (3.94,-2.37)--(7.51,-0.64) node {\color{black}.} node[above] {\color{black}{3}}
--(7.80,-0.49);

\draw[purple] (4.30,-0.97)--(4.77,-1.35) node {\color{black}.} node[above] {\color{black}{2}}
--(7.36,-3.51);

\draw[gray] (5.26,-0.38)--(5.26,-0.62) node {\color{black}.} node[left] {\color{black}{1}}
--(5.26,-3.58);

\draw[green] (3.46,-0.78)--(6.67,-2.50) node {\color{black}.} node[below] {\color{black}{5}}
--(7.54,-2.95);

\node at (5.26,-1.73) {.} ;

\node at (5.13,-2.04) {7};

\end{scope}

\end{tikzpicture}
\caption{A configuration of singular solutions in $X(4,7)$.  \textit{Left}: Near the soft limit  four 2-planes $\overline{123}$, $\overline{345}$, $\overline{561}$ and $\overline{246}$ almost cross the soft particle.  \textit{Right}: In the strict soft limit the soft particle lies in the intersection of the four 2-planes.}
\label{x47t}
\end{figure}
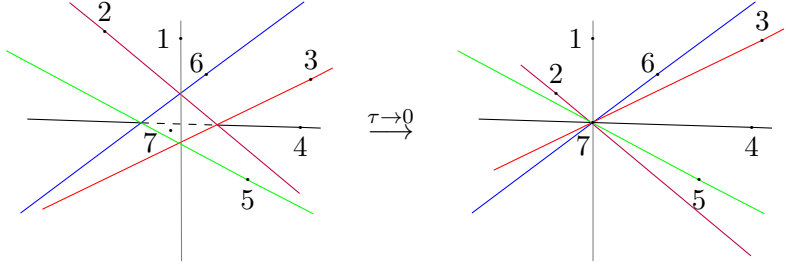

Another simple example is $X(4,7)$. Taking again particle $7$ to be soft, {\it i.e.} $s_{abc7}\to \tau {\hat s}_{abc7}$ (with $\tau \to 0$), all  singular solutions come from 30 different configurations where determinants of the form $|1237|$, $|3457|$, $|1567|$ and $|2467|$ vanish.  
We can geometrically interpret this configuration  in the soft limit as
having the soft particle as the intersection of four planes, and each hard particle lying on the intersection of two of those planes. We give a very schematic representation, i.e. drawing $\mathbb{RP}^3$ on a plane, of this in figure \ref{x47t}.
There are $4$ solutions for each of these configurations, so we obtain a total amount of ${\cal N}_7^{(4):{\rm singular}}=4\times 30=120$ singular solutions. 

The way  to get the solutions is the following.
For a configuration where $|1237|$, $|3457|$, $|1567|$ and $|2467|$ vanish,  a convenient choice of gauge fixing in projective space is
\be
\left(
                    \begin{array}{ccccccc}
                      1 & 0 & 1 &0   &0   & 1 & 1 \\
                      x_{1} & x_{2} & x_{3}   & x_{4} &x_{5}   &x_{6}  &x_{7}  \\
                      y_{1} & y_{2} & y_{3}  & y_{4}    &y_{5}   &y_{6}  &y_{7}   \\
                       z_{1} & z_{2} & z_{3}  & z_{4}    &z_{5}   &z_{6}  &z_{7}   \\
                    \end{array}
                  \right)\overset{\text{gauge~fixing}}{\to}
\left(
                    \begin{array}{ccccccc}
                      1 & 0 & 1 &0    &0   & 1 & 1 \\
                      0 & 1 & 1   & 0 &0   &x_{6}  &x_{7}  \\
                      0 & 0 & 1  &1    &0   &y_{6}  &y_{7}   \\
                        0 & 0 & 1  &0  &1   &z_{6}  &z_{7}   \\
                    \end{array}
                  \right).
\ee
We then parameterize each determinant as $
|1237|=\tau u$, $\quad |3457|=\tau  v$, $\quad |5617|=\tau  p$, and $\quad |2467|=\tau  q$, that is
\be\label{para4}
x_6= \frac{\tau  (v y_6+ p)+y_6}{ \tau (q-u) +z_6},\quad x_7= \tau  v+1, \quad y_7=  \tau (q-  u)+z_6,\quad z_7=
   \tau q +z_6\,.
\ee
When we plug this into the original scattering equations and take the strict soft limit $\tau\to 0$, we obtain a new set of scattering equations with variables $u$, $v$, $p$, $q$, $y_{6}$ and $z_{6}$
\be\label{neweq4}
\left. \lim_{\tau\to0} \,  \frac{\partial {\cal S}_{4}}{\partial x_{i}}\right| _{\eqref{para4}}=0, \quad  \lim_{\tau\to0} \, \left.  \frac{\partial {\cal S}_{4}}{\partial y_{i}}\right|_{\eqref{para4}}=0,  \quad  \lim_{\tau\to0} \, \left.  \frac{\partial {\cal S}_{4}}{\partial z_{i}}\right| _{\eqref{para4}}=0, \quad \text{for ~} i=1,\ldots 7.
\ee
Among the above 21 equations, only 6 of them are independent. The system is simple enough that all variables except one can be eliminated using resultants producing an irreducible polynomial of degree $4$ for the left over variable. This means that there are $4$ solutions.  
Note that in this case one can again eliminate $u$, $v$, $p$ and $q$ in the new scattering equations \eqref{neweq4} first and then reduce the system to one only involving hard particles
\be
\left. \Big(\frac{\partial \tilde{\cal S}_{4}}{\partial x_{6}}+ z_{6}\frac{\partial \tilde{\cal S}_{4}}{\partial y_{6}}\Big)\right|_{x_{6}\to \frac{y_{6}}{z_{6}}}=0, \quad \left. \Big(\frac{\partial \tilde{\cal S}_{4}}{\partial z_{6}}+x_{6} \frac{\partial \tilde{\cal S}_{4}}{\partial y_{6}}\Big)\right|_{x_{6}\to \frac{y_{6}}{z_{6}}}=0,
\ee
where ${\tilde{\cal S}_{4}}$ is the potential of hard particles,
$
{\tilde{\cal S}_{4}}\equiv \sum_{1\leq a<b<c<d\leq 6}s_{abcd}\log |a, b , c,d|
$. Here we just present two independent equations and the remaining variables are $y_{6}$ and $z_{6}$.  Compared to the original scattering equations for 6 particles, which has 6 solutions, now the requirement that the planes $\overline{123}$, $\overline{345}$, $\overline{561}$  and $\overline{246}$ pass through a common point reduces the number of solutions to 4. Therefore, the number of singular solutions for $X(4,7)$ is ${\cal N}_7^{(4):{\rm  singular}}=30\times 4=120$, as expected.

\subsubsection{Singular Solutions on Positive Kinematics}

Now, we make the use of kinematic data in the positive region ${\cal K}_{4,n}^+$ (or more precisely in ${\cal K}_{4,n}^{+,\mathbb{R}}$) \eqref{potential4} to study the solutions. 
Singular solutions make some minors of the form $|abc7|$, i.e. containing particle 7, vanish. Geometrically, this means that planes  $\overline{abc}$ in  ${\mathbb {RP}}^{3}$ space  will dominate. The remaining planes can be omitted for the soft particle at first. Therefore, in order to bound particle $7$ in  ${\mathbb {RP}}^{3}$ space, we need at least $4$ such dominating planes. That is, we need at least four vanishing minors involving particle $7$ while keeping the other minors still finite. For $n=7$, this can be achieved for example by letting $|1237|$, $|3457|$, $|1567|$ and $|2467|$ vanish. In figure \ref{x47t} the soft particle is bounded inside a tetrahedron whose volume is of order $\tau$ when we use the kinematic data from the positive region.  It is not trivial for six hard particles to form such a tetrahedron while any four of them are not allowed to lie in a common plane. One can check that this is the only kind of configuration, up to relabelling, that achieves this goal for $X(4,7)$.   

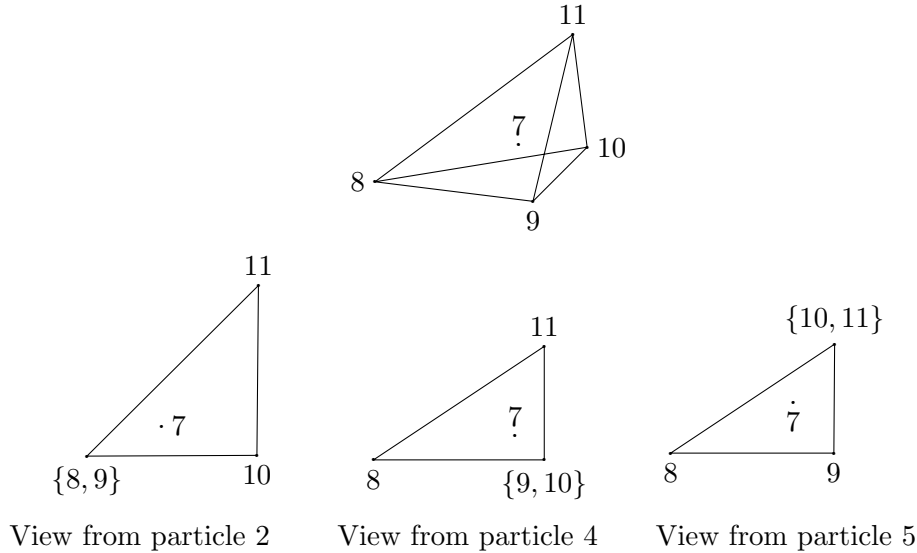
\begin{figure}[!htb]
\centering
\begin{tikzpicture} [xscale=1,yscale=1]

\draw (6.04,-3.62)node {.} node[left] {8}--
(8.14,-3.88) 
node {.}  node[below] {9}
--(8.86,-3.16)node {.}  node[right] {10}
--(8.67,-1.66)node {.}  node[above] {11}
--(6.04,-3.62)
(6.04,-3.62)--(8.86,-3.16)
(8.14,-3.88) --(8.67,-1.66)

(2.21,-7.27)node{.}node[below]{$\{8,9\}$}
--(4.47,-7.26)node{.}node[below]{10}
--(4.49,-5.00)node{.}node[above]{11}--
(2.21,-7.27)

(6.02,-7.32)node{.}node[below]{8}
--(8.29,-7.32)node{.}node[below]{$\{9,10\}$}
--(8.29,-5.81)node{.}node[above]{11}--
(6.02,-7.32)

(9.97,-7.23)node{.}node[below]{8}
--(12.14,-7.23)node{.}node[below]{9}
--(12.15,-5.78)node{.}node[above]{$\{10,11\}$}--
(9.97,-7.23)

(7.95,-3.13) --(7.95,-3.13)  node{.}node[above]{7}

(3.20,-6.87) --(3.20,-6.87)  node{.}node[right]{7}

(7.90,-7.00) --(7.90,-7.00) node{.}node[above]{7}

(11.60,-6.55) --(11.60,-6.55) node{.}node[below]{7}

;

\end{tikzpicture}

View from particle 2 \qquad View from particle 4\qquad View from particle 5
\caption{\textit{Top}: The soft particle lies inside a tetrahedron. \textit{Bottom}: Three projections of the tetrahedron from the point of view of particles 2, 4 and 5, respectively, when these are sent to infinity. In the strict soft limit, the tetrahedron as well as its three projections collapse to a point. }
\label{47f4}
\end{figure}

Here we introduce another description of the tetrahedron, which can be generalized to describe more complicated polytopes. 
We view  each vertex of the  tetrahedron as an auxiliary point and give each of them  a label, ranging from 8 to 11, see figure \ref{47f4}.  
 
Note that particle labels are from 1 to 7.  Hard particles 1-6 lie on the lines determined by \{8,11\}, \{8,9\}, \{9,11\}, \{9,10\}, \{10,11\} and \{8,10\} respectively. 
Using the auxiliary points, we can understand the relative positions of the  hard particles. Alternatively, now we can ignore the auxiliary points and imagine how these hard particles form some dominating planes to bound the soft particle.

We can also describe this tetrahedron through its projections from 3 orthogonal directions. As particles 2, 4 and 5 are sent to infinity in different directions, we can say that the three projections in figure \ref{47f4} are just what the tetrahedron would look like if one stands at the position of  2, 4 and 5 respectively. In the first projection, vertices 8 and 9 are pinched from the view of particle 2, which has been sent to infinity. We can say that particle 2 lies on the lines determined by \{8,9\}. The remaining two projections are completely analogous.

There are 4 solutions for this particular configuration. For generic points in
${\cal K}_{4,7}^{+}$ there are complex solutions. However, the region ${\cal K}_{4,7}^{+,\mathbb{R}}$ is non-empty and therefore resting to it one can find all four real solutions. Looking at the new equations \eqref{neweq4} for $X(4,7)$, it is very hard to see whether there are solutions. Using the positive kinematic data in ${\cal K}_{4,7}^{+,\mathbb{R}}$ and viewing the solutions as equilibrium points, we see at least that there are possible solutions for the soft particle 7. 


A beautiful way to count the number of solutions is from the dual limit in the dual space $X(3,7)$ as we explain now. 

\subsubsection{Singular Solutions from a Dual Hard Limit}

One new feature of $k>2$ kinematics is that in addition to soft limits there are also ``hard limits''. In fact, these are dual to each other under the isomorphism $X(k,n)\sim X(n-k,n)$ with the corresponding action on kinematic invariants \cite{CEGM,GG}. In the case at hand, the soft limit of particle $7$ in $X(4,7)$ is dual to the hard limit of particle $7$ in $X(3,7)$. It is important not to confuse it with the soft limit of particle $7$ in $X(3,7)$ analyzed at the beginning of the section. 

The reason for the name is easily seen from the relation among kinematic invariants. Consider  $X(4,7)\sim X(3,7)$ and the relation $s_{abcd}=s_{efg}$ with $\{e,f,g\}=\{1,2,\ldots ,7\}\setminus \{a,b,c,d\}$. This means that the soft limit in $X(4,7)$, i.e. $s_{abc7}\to 0$ with the rest finite, implies $s_{abc}\to 0$ if $7\notin{\{a,b,c\}}$ and finite for any invariant containing $7$, i.e. $s_{ab7}$. 
Going back to the singular solutions in $X(4,7)$, one can explicitly visualize the four solutions for each of the $30$ configurations by using the dual hard limit in $X(3,7)$. Recall that the singular solutions in $X(4,7)$ come from configurations where the determinants of the form $|1237|$, $|3457|$, $|1567|$ and $|2467|$ vanish. Just as for kinematic invariants, this corresponds to having the determinants $|456|$, $|126|$, $|234|$ and $|135|$ vanishing in $X(3,7)$.

If we gauge fix the homogeneous coordinates of particles $3$ and $6$ to infinity as $(0,0,1)$ and $(0,1,0)$, and of particles $4$ and $1$ to be the origin $(1,0,0)$ and $(1,1,1)$, then the configurations that give rise to singular solutions automatically fix particles $2$ and $5$ to be at $(1,0,1)$ and $(1,1,0)$ respectively. 

\begin{figure}[!htb]
\centering
\includegraphics[width=130mm]{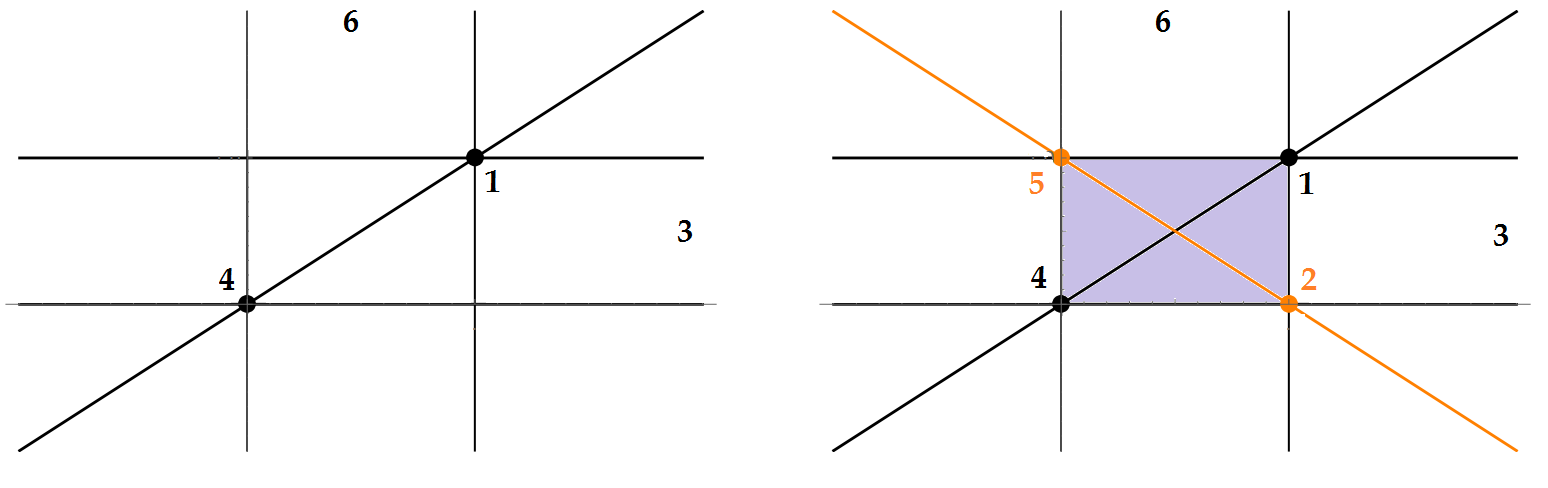}
\caption{Four bounded chambers from the hard limit in $X(4,7)$. \textit{Left:} the gauge-fixed particles 1, 3, 4 and 6 create repelling black lines. \textit{Right:} the singular configurations automatically fix the position of particles 2 and 5 to be in the two remaining vertices of the square $[0,1]^2$, and create a new repelling (orange) line. This produces four bounded chambers (shown in grey) where particle 7 can be.}
\label{x47}
\end{figure}

Therefore, for generic positive kinematics in $\mathbb{RP}^2$ we are left with four bounded chambers, which correspond to equilibrium points where particle $7$ can be. These points correspond to the $4$ solutions of the system.  
We give a graphical representation in figure \ref{x47}.

\section{Singular Solutions in $X(3,8)\to X(3,7)$ and $X(5,8)\to X(5,7)$}\label{sec5}

Now we move on to two more complicated cases, $X(3,8)$ and $X(5,8)$, each of which having their own new features. In the former, the equations are complicated enough that counting solutions directly is not straightforward as in previous cases. Instead we use that the new scattering equations at $\tau=0$ can also be analysed in soft limits to count solutions. The new equations also turn out to have both regular and singular solutions. In the latter case, we find the first example in which several topologically distinct configurations contribute to singular solutions. Of course, we expect this to be the generic behavior for higher $k$ and $n$.

\subsection{Singular Solutions in $X(3,8)\to X(3,7)$} \label{sec51}

In order to obtain the singular solutions, we study the soft limit for, e.g., particle $n=8$, {\it i.e.}  $s_{ab8}\to\tau \hat{s}_{ab8}$ (with $\tau\to 0$).  The singular solutions arise from configurations where three lines in $\mathbb{CP}^2$, each defined by two hard particles, meet at the soft particle. In the same spirit as for $X(3,7)$, we let e.g. $|148|$, $|258|$ and $|368|$ vanish and we find $568$ singular solutions. The large number of the solutions is the reason this case resists a direct approach as mentioned above.
There are ${7 \choose 2}{5 \choose 2}{3 \choose 2}/3!=105$ different configurations of this kind and therefore ${\cal N}_8^{(3):{\rm  singular}}=105\times 568 = 59\,640$.

Let us now explain how to count the solutions for each singular configuration. The $568$ solutions can be counted by taking a second soft limit, say that of particle $7$. The solutions to the new scattering equations come in three different types. The first corresponds to regular solutions and the other two to singular solutions. Since the three kinds come from  the singular solutions for the particle $8$  we can denote them as $({\texttt {regular}}_7,{\texttt {singular}}_8)$ and $({\texttt {singular}}_7,{\texttt {singular}}_8)$ of type $A$ and type $B$.

The first class of solutions, $({\texttt {regular}}_7,{\texttt {singular}}_8)$, come from decoupling particle $7$ from the remaining hard particles. We obtain $12$ solutions for the hard particles and each gives $41$ solutions for particle $7$ leading to $12\times41=492$ such solutions. 

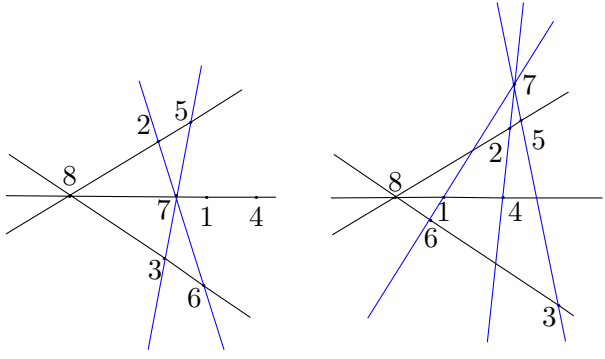
\begin{figure}[!htb]
\centering
\begin{tikzpicture} [xscale=1,yscale=1]

\draw (4.61,-2.72)--(7.28,-2.74) node {.} node[below] {1}
--(7.94,-2.74)node {.}  node[below] {4}
--(8.20,-2.74)

(4.61,-3.23)--(6.64,-2.00)node{.}node[above left=-1pt]{2}--(7.07,-1.74)node{.}node[above left=-3pt]{5}--
(7.75,-1.31)

(4.65,-2.17)--(6.72,-3.55)node{.}node[below left=-3pt]{3}--(7.24,-3.92)node{.}node[below left=-3pt]{6}--(7.86,-4.34);

\draw[blue](6.38,-1.19)--(7.51,-4.77)

(7.21,-0.99)--(6.50,-4.77)
;

\node at (5.46,-2.73) {.} ;

\node at (5.46,-2.46) {8};

\node at (6.88,-2.72) {.} ;

\node at (6.71,-2.90) {7};

\draw(8.93,-2.75) -- (10.43,-2.74)node {.} node[below=-2pt] {1}
--(11.22,-2.75) node {.}  node[below right=-2pt] {4}
--(12.54,-2.75)

(8.95,-3.24)--(11.31,-1.82)node{.}--(11.46,-1.72)node{.}node[below right=0pt]{5}--
(12.09,-1.34)

(8.97,-2.17)--(10.26,-3.05)node{.}node[below=-1pt]{6}--(11.96,-4.18)node{.}node[below left=-3pt]{3}--(12.16,-4.32);

\node at (11.12,-2.15) {2};
\draw[blue](9.42,-4.35)--(11.89,-0.40)

(11.01,-4.67)--(11.49,-0.15)

(12.05,-4.66)--(11.14,-0.15)
;

\node at (9.79,-2.74) {.} ;

\node at (9.79,-2.53) {8};

\node at (11.37,-1.24) {.} ;

\node at (11.57,-1.24) {7};

\end{tikzpicture}
\caption{\textit{Left}: representation of $({\texttt {singular}}_7,{\texttt {singular}}_8)$ type A configurations. \textit{Right}: representation of $({\texttt {singular}}_7,{\texttt {singular}}_8)$ type B configurations. \label{38twof} }
\end{figure}

The first kind of $({\texttt {singular}}_7,{\texttt {singular}}_8)$ solutions come from configurations in which particle $7$ belongs to one of the already existing three lines and lies in the intersection of two other new lines (see left side of figure \ref{38twof}). For instance, this would correspond to vanishing determinants of the form $|147|$, $|267|$ and $|357|$. We find $6$ solutions for each of the $6$ possible configurations of this kind. 

The second kind of $({\texttt {singular}}_7,{\texttt {singular}}_8)$ solutions corresponds to the case where three
new lines intersect at particle $7$ (see right side of figure \ref{38twof}). For instance, this would correspond to vanishing determinants of the form $|167|$, $|247|$ and $|357|$. We find $5$ solutions for each of the $8$ possible configurations of this kind. 

Notice that in these last two cases there is a symmetry between particles 7 and 8. Combining these results one finds that the number of solutions to the equations that arise in a particular singular configuration in the soft limit of particle $8$ is $41\times12+8\times5+6\times6=568$.

Let us now explain how the procedure is implemented. We can again use a similar gauge fixing as in previous cases and parameterize the space as
\be\label{para38}
  x_{6} = y_{4} - \tau ( u  + v + p),\quad
x_{8} = y_{4} -  \tau ( u + v) ,  \quad y_{8} = y_{4} -\tau v.
\ee
When we take the strict soft limit $\tau\to 0$ we obtain a set of new equations with $8$ variables: $u$, $v$, $p$, $x_{5}$, $y_{5}$, $y_{6}$, $x_{7}$ and $y_{7}$. The equations are given by
\be\label{nneweq38}
\left. \lim_{\tau\to0} \,  \frac{\partial {\cal S}_{3}}{\partial x_{i}}\right| _{\eqref{para38}}=0, \quad \left.  \lim_{\tau\to0} \,  \frac{\partial {\cal S}_{3}}{\partial y_{i}}\right| _{\eqref{para38}}=0,  \quad \text{for ~} i=1,\ldots 8.
\ee
More explicitly, we have
\begin{align}
\left. \lim_{\tau\to0} \,  \frac{\partial {\cal S}_{3}}{\partial x_{8}}\right|_{\eqref{para38}} &= \frac{s_{368}}{p}-\frac{s_{148}}{u}\,,\quad
\left. \lim_{\tau\to0} \,  \frac{\partial {\cal S}_{3}}{\partial x_{8}}\right| _{\eqref{para38}}=\frac{s_{148}}{u}-\frac{s_{258}}{v}\,,
\nonumber
\\
\left. \lim_{\tau\to0} \,  \frac{\partial {\cal S}_{3}}{\partial y_{5}}\right| _{\eqref{para38}}&= \frac{s_{258}}{v} +  \left. \frac{\partial {\tilde {\cal S}}_{3}}{\partial y_{5}}\right|_{x_{6}\to y_{5}}\,,
\quad
\left. \lim_{\tau\to0} \,  \frac{\partial {\cal S}_{3}}{\partial x_{6}}\right| _{\eqref{para38}}= -\frac{s_{368}}{p} +  \left. \frac{\partial {\tilde {\cal S}}_{3}}{\partial y_{5}} \right| _{x_{6}\to y_{5}}\,,
\end{align}
with
${\tilde{\cal S}_{3}}$ defined as the potential ${\tilde{\cal S}_{3}}\equiv \sum_{1\leq a<b<c\leq 7}s_{abc}\log (a, b , c)
$. This allows us to easily eliminate $u$, $v$ and $p$ and reduce the new set of equations \eqref{nneweq38} to ones involving less particles,
\be\label{neweq38}
\left. \left\{\frac{\partial \tilde{\cal S}_{3}}{\partial y_{5}}+ \frac{\partial \tilde{\cal S}_{3}}{\partial x_{6}}\,,
\quad  \frac{\partial \tilde{\cal S}_{3}}{\partial x_{5}}\,,
 \quad  \frac{\partial \tilde{\cal S}_{3}}{\partial y_{6}}\,,
 \quad
\frac{\partial \tilde{\cal S}_{3}}{\partial x_{7}}\,,
  \quad  \frac{\partial \tilde{\cal S}_{3}}{\partial y_{7}}  \right\}
\right|_{x_{6}\to y_{5}}=0\,.
\ee
Now we have 5 independent equations and the remaining variables are $x_{5}$, $y_{5}$, $y_{6}$, $x_{7}$ and $y_{7}$. So we have reduced a problem of $8$ particles to one of $7$ particles. 

This means we can now start with the set of equations \eqref{neweq38} and study them independently of where they came from, i.e. the eight-particle problem, just like we did in the $X(3,7)$ case. As mentioned above, when the soft limit $s_{ab7}\to \epsilon {\hat s}_{ab7}$ (with $\epsilon \to 0$) is taken, there are again both regular and singular solutions for \eqref{neweq38}.

For the $({\texttt {regular}}_7,{\texttt {singular}}_8)$ solutions, all terms involving particle $7$ can be omitted in the first three equations in \eqref{neweq38},  which become exactly the same equations as \eqref{neweq37} and give $12$ solutions for $x_{5}$, $y_{5}$ and $y_{6}$. We plug each solution into the last two equations in \eqref{neweq38} and obtain $41$ solutions for $x_{7}$ and $y_{7}$. Compared to the regular solutions in $X(3,7)$, roughly speaking, we can see that the requirement $x_{6}=y_{5}$ reduces the number of solutions for $x_{7}$ and $y_{7}$ from $42$ to $41$. In total, we obtain $12\times 41=492$ solutions from this sector. We give a graphical representation and counting of these solutions in appendix \ref{appendix22}.

For the $({\texttt {singular}}_7,{\texttt {singular}}_8)$ solutions, note that the three lines $\overline{12}$, $\overline{34}$ and $\overline{56}$ already intersect at the same point (i.e. the position of particle 8, but this fact is irrelevant for our present problem). Now we need another $3$ lines intersecting at the position of particle 7 in the ${\mathbb {CP}}^{2}$ in the strict second soft limit $\epsilon\to 0$. There are $2$ different kinds of configurations that we graphically represented in figure \ref{38twof}.

For the first configuration, which corresponds to requiring $|147|$, $|267|$ and $|357|$ to vanish, we re-parameterize the ${\mathbb {CP}}^{2}$ space using the constraints
\be
|147|=\epsilon\, q, \quad |267|=\epsilon \, r, \quad |357|=\epsilon\,  s,
\ee
that is
\be\label{para45}
y_6=  \epsilon(q+r+s)
   +x_5,
  \quad
   x_7= s \epsilon +x_5,
   \quad
   y_7=  \epsilon(q +s ) +x_5.
\ee
If one plugs them in into \eqref{neweq38} and takes the strict soft limit $\epsilon\to 0$, a new set of scattering equations with variables $q$, $r$, $s$, $x_{5}$ and $y_{5}$ arises. Again we can easily eliminate $q$, $r$ and $s$ and reduce the new set of equations to those only involving the six hard particles
\be\label{neweq381}
\left. \frac{\partial \hat{\cal S}_{3}}{\partial y_{5}}+ \frac{\partial \hat{\cal S}_{3}}{\partial x_{6}}\right| _{x_{6}\to y_{5}, y_{6}\to x_{5}}=0\,, \left.
\quad  \frac{\partial \hat{\cal S}_{3}}{\partial x_{5}}+
  \frac{\partial \hat{\cal S}_{3}}{\partial y_{6}}
\right|_{x_{6}\to y_{5}, y_{6}\to x_{5}}=0\,,
\ee
with
${\hat{\cal S}_{3}}$ defined as the potential of $6$ particles, i.e.
$
{\hat{\cal S}_{3}}\equiv \sum_{1\leq a<b<c\leq 6}s_{abc}\log (a, b , c)
$.
It turns out that there are 6 solutions to these equations.

For the second type of configuration, i.e., where $|167|$, $|247|$ and $|357|$ are taken to vanish, we re-parameterize the ${\mathbb {CP}}^{2}$ space using the constraints
\be
|167|=\epsilon\, q, \quad |247|=\epsilon \, r, \quad |357|=\epsilon\,  s,
\ee
that is
\be\label{para452}
x_7=  \epsilon s +x_5,\quad y_7= 1- \epsilon r, \quad y_6= \frac{- \epsilon(q+ r y_5 ) +y_5}{
   \epsilon s+x_5}.
\ee
When we plug this into \eqref{neweq38} and take the strict soft limit $\epsilon\to 0$, we obtain a set of new scattering equations with variables $q$, $r$, $s$, $x_{5}$ and $y_{5}$. Again we can easily eliminate $q$, $r$ and $s$ and reduce the new set of equations to those only involving $6$ hard particles
\be\label{neweq382}
\left. \frac{\partial \hat{\cal S}_{3}}{\partial y_{5}}+ \frac{\partial \hat{\cal S}_{3}}{\partial x_{6}} + \frac{x_{5}} {y_{5}} \frac{\partial \hat{\cal S}_{3}}{\partial x_{5}}\right|_{x_{6}\to y_{5}, y_{6}\to x_{5}}=0 \,, \left.
\quad   \frac{x_{5}^{2}} {y_{5}}   \frac{\partial \hat{\cal S}_{3}}{\partial x_{5}}+
  \frac{\partial \hat{\cal S}_{3}}{\partial y_{6}}
\right|_{x_{6}\to y_{5}, y_{6}\to x_{5}}=0.
\ee
These equations have $5$ solutions. Therefore, we obtain a total of $6\times 6+8\times 5=76$ $({\texttt {singular}}_7,{\texttt {singular}}_8)$ solutions.

Summarizing, we have proven that there are $568$ solutions for the new set of equations \eqref{neweq38}. Therefore, as mentioned above, we find that the total number of singular solutions for $X(3,8)$ is ${\cal N}_8^{(3):{\rm  singular}}=105\times 568=59\, 640$. Together with the already known $128\, 472$  regular solutions, mentioned in section \ref{sec3}, we get a total of ${\cal N}_8^{(3):{\rm total}}=188\,112$ solutions, which is consistent to a proposal made by Lam that this should be related to the number of representations of uniform matroids as defined in \cite{skorobogatov1996number}.

Of course, the challenge now is to reproduce the same number of total solutions for the dual space $X(5,8)$. Confirming that ${\cal N}_8^{(5):{\rm total}}=188\,112$ would be a very strong consistency check on our constructions and on the number itself.

\subsection{Singular Solutions in $X(5,8)\to X(5,7)$}

Unlike any case considered previously, the soft limit $X(5,8)\to X(5,7)$ has four kinds of topologically distinct singular solutions. In order to describe them let particle $8$ be soft. As summarized in table \ref{458}, the four kinds of  singular solutions come from the configurations where either $4$, $5$, $6$, or $7$ minors involving particle 8 vanish. Each class has $210$, $420$, $210$, and $840$ different configurations, respectively. For each of them, there are $96$, $24$, $8$, and $32$ solutions respectively, as we show below. Thus we obtain ${\cal N}_8^{(5):{\rm  singular}}=210\times 96+420\times 24+ 210\times8+ 840\times32 =58\,800 $ singular solutions. 

\begin{table}[!htb]
\centering
\begin{tabular}{c|c|c|c}
 ${\rm Topology \atop type}$& Vanishing minors& ${\rm Number~ of \atop configurations}$  &  ${\rm Number~ of \atop solutions}$ \\
 \hline
1&$ |5 7 2 38|, |5 7 1 48|, |5 3 4 68|, |5 1 2 68|$ & 210  &  96 \\
2& $|1 2 3 5 8|,|1 2 4 6 8|,|156 78|,|2 3 4 78|,|3 4 5 68|$&  420 & 24   \\
3&
$|1 2 3 78|, |1 2 4 58|, |1 3 5 68|, |2 3 4 68|, |2 5 6
  78|, |3 4 5 78|$ & 210  &  8 \\
  4&
 $|1 2 3 48|,|1 2 3 58|,|1 2 3 68|,|1 2 3 78|,|1 4 5
  68|,|2 4 5 78|,|3 4 6 78|$ &  840&32
\end{tabular}
\caption{Four different topologies of singular configurations for $X(5,8)$ defined by the list of vanishing minors.}
\label{458}
\end{table}

\subsubsection{Type 1 Configuration } \label{vanishleading}
\begin{figure}[!htb]
\centering
\includegraphics[width=.4\textwidth]{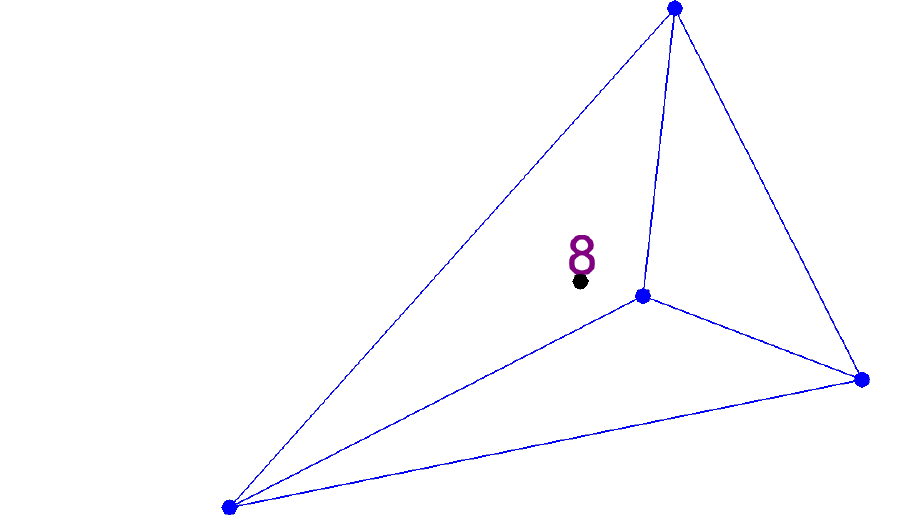}
\caption{The geometrical interpretation of the topology type 1 near the soft limit, from the point of view of particle 5, is that the projection of the soft particle lies inside a tetrahedron just like in $X(4,7)$. In the strict soft limit, in projection space, the tetrahedron collapses to a point where the projection of the soft particle 8 lies while in ${\mathbb{CP}}^4$ space,  the four  3-planes ${\overline {5723}}$,  ${\overline {5714}}$, ${\overline {5346}}$, and  ${\overline {5128}}$ share a common line  that crosses particles 5 and 8.}
\label{11fourf4}
\end{figure}

The first configuration in table \ref{458} is slightly subtler than the remaining ones so we describe it in a separate subsection.  The soft particle 8 can be thought of as being projected from ${\mathbb{CP}}^{4}$ space to ${\mathbb{CP}}^{3}$ space through the particle 5 and then being crossed by four 2-planes in the projection space just like the case $X(4,7)$.    We give a very schematic representation, i.e. drawing ${\mathbb {RP}}^{3}$ on a plane, of the projection in figure \ref{11fourf4}, where particle 5 is sent to infinity. 
 The four vertices of the tetrahedron in this projection actually correspond to four parallel lines in  $\mathbb {RP}^4$ space. We can think of these four lines intersecting at the infinity point, particle 5. 

There are no analogs to this case for $k=3$ because there are no singular solutions for $k=2$.  Starting at $k=4$, however, the soft particle can be projected into a lower-dimensional space and  its projection satisfies the requirement of that particular dimension as long as $n$ is large enough.

The way to obtain the solutions in the soft limit $s_{abcd8}\to \tau {\hat s}_{abcd8}$ (with $\tau \to 0$) is the following.  First, we parameterize $X(5,8)$ as
\be\label{cp4}
\left(
                    \begin{array}{cccccccc}
                      1 & 0 & 0 &0    &0   & 1 & 1& 1 \\
                      x_{1} & x_{2} & x_{3}   & x_{4} &x_{5}   &x_{6}  &x_{7} &x_{8}  \\
                      y_{1} & y_{2} & y_{3}  & y_{4}    &y_{5}   &y_{6}  &y_{7}  &y_{8}  \\
                                            z_{1} & z_{2} & z_{3}   & z_{4} &z_{5}   &z_{6}  &z_{7} &z_{8}  \\
                      w_{1} & w_{2} & w_{3}  & w_{4}    &w_{5}   &w_{6}  &w_{7}  &w_{8}  \\
                    \end{array}
                  \right)\overset{\text{gauge~fixing}}{\longrightarrow}
\left(
                   \begin{array}{cccccccc}
                      1 & 0 & 0 &0    &0   & 1 & 1& 1 \\
                     0 & 1 & 0   & 0 &0   &1  &x_{7} &x_{8}  \\
                     0 & 0 & 1  & 0    &0  &1  &y_{7}  &y_{8}  \\
                      0 &0 & 0   & 1 &0  &1 &z_{7} &z_{8}  \\
              0 &0 & 0  & 0    &1  &1  &w_{7}  &w_{8}  \\
                    \end{array}
                  \right).
\ee
Notice that a direct consequence of sending particle 5 to infinity in the direction of $w$ is that all vanishing minors 
$|5723 8|$, $|5714 8|$, $|53468|$ and $|5126 8|$ become independent of $w_8$. 

Next, we make a reparameterization under the constraints
\be
|5723 8|=\tau u,\,\, |5714 8|=\tau v, \,\,|53468|=\tau p,\,\,|5126 8|=\tau q\,,
\ee
that is
\begin{align}\label{para584}
y_7= \frac{\tau(-q   x_7+  u x_7-v)+x_7
   z_7}{\tau  p+1},\,\,\,
   x_8= \tau  p+1,
   \,\,\,y_8=  \tau(
   u-q )+z_7,\,\,\,z_8= \tau  u+z_7.
\end{align}
We then plug this into the original scattering equations and take the strict soft limit $\tau\to 0$
\be\label{neweq584}
\left. \left\{\lim_{\tau\to0} \, \frac{\partial {\cal S}_{5}}{\partial x_{i}} \,,\quad
\lim_{\tau\to0} \,\frac{\partial {\cal S}_{5}}{\partial y_{i}} \,,\quad
\lim_{\tau\to0} \,\frac{\partial {\cal S}_{5}}{\partial z_{i}} \,,\quad
\lim_{\tau\to0} \,\frac{\partial {\cal S}_{5}}{\partial w_{i}}\right\}
\right|_{\eqref{para584}}=0,  \quad \text{for ~} i=1,\ldots 8.
\ee
Since all vanishing minors are independent of $w_8$, 
the above $32$ equations only depend on $7$ variables $u$, $v$, $p$, $q$, $x_{7}$, $z_{7}$ and $w_{7}$. Correspondingly, only $7$ of these equations are independent since e.g. the leading order of $\,\frac{\partial {\cal S}_{5}}{\partial w_{8}}$ in $\tau$ vanishes.  Hence, we must require its subleading contribution to vanish
\be\label{sub}
\left. \lim_{\tau\to0} \,\frac{1}{\tau}\frac{\partial {\cal S}_{5}}{ \partial w_{8}} \right| _{\eqref{para584}} =0.
\ee
One can solve the equations for the leading order in \eqref{neweq584} and obtain $16$ solutions for $u$, $v$, $p$, $q$, $x_{7}$, $z_{7}$ and $w_{7}$. When each of these solutions is plugged into the subleading term \eqref{sub}, we find $6$ solutions for $w_{8}$. Therefore, the total number of solutions is $16\times 6=96$ as shown in table \ref{458}. 

Now we can again use the kinematic data from the positive region \eqref{potential4}, assuming all solutions are real, to interpret the singular solutions.  
Singular solutions  make some minors $|abcd8|$ containing particle 8  to become singular.  Geometrically, this means that 3-planes  $\overline{abcd}$ in  ${\mathbb {RP}}^{4}$ space  will dominate. The remaining planes can be omitted for the soft particle at first. Therefore, in order to bound particle $8$ in  ${\mathbb {RP}}^{4}$ space, it seems we need at least five such dominating planes. 
This is the case for the remaining 3 configurations in table \ref{458}. However, it is not the case for type 1 configuration described now. 
The four dominating 3-planes don't bound the soft particle. They produce equilibrium lines instead of equilibrium points  for the soft particle.

As shown in figure 
\ref{11fourf4},  any equilibrium point in the projection space will correspond to an equilibrium line in ${\mathbb {RP}}^{4}$ space.  The soft particle 8 can lie at any point of these equilibrium lines  and won't be pushed to infinity by the dominating 3-planes.    This corresponds to the fact that the leading order of $\,\frac{\partial {\cal S}_{5}}{\partial w_{8}}$ in $\tau$ vanishes.  It has no constraints on the positions of particle  8 in the direction from which 5 is sent to infinity.  

The position of the soft particle 8 is finally determined by considering the normal 3-planes determined by the hard particles as well.   This corresponds to equation \eqref{sub}.
In each of the equilibrium lines, 
there are 6 equilibrium points considering both dominating and non-dominating minors.

\subsubsection{Other Three Types of Configurations}

For the second configuration in table \ref{458}, we make a reparameterization of \eqref{cp4} under the constraints
\be
|1 2 3 5 8|=\tau u,\,\, |1 2 4 6 8|=\tau v, \,\,|156 78|=\tau p,\,\,|2 3 4 78|=\tau q, \,\,|3 4 5 68|=\tau r\,,
\ee
that is
\begin{align}\label{para581}
  & w_7= -\frac{\tau( -r   y_7+r   z_7-q
   x_7+q   z_7+  u x_7-  u y_7+  v x_7-  v z_7+
   p)-y_7+z_7}{x_7-z_7}, \nonumber
\\
&
x_8=  \tau r+1,\quad
y_8= -\frac{\tau (r   z_7+  u x_7-r  y_7- u y_7+
   p)-y_7+z_7}{x_7-z_7},
   \quad
    \quad  z_8= -\tau  u,\nonumber
    \\&
    w_8= -\frac{\tau(-r   y_7+r   z_7+  u x_7-  u
   y_7+  v x_7-  v z_7+  p)-y_7+z_7}{x_7-z_7}\,.
\end{align}
When this is plugged into the original scattering equations and the strict soft limit $\tau\to 0$ is taken, we obtain a new set of scattering equations with variables $u$, $v$, $p$, $q$, $y_{6}$ and $z_{6}$
\be\label{neweq581}
\left. \left\{\lim_{\tau\to0} \, \frac{\partial {\cal S}_{5}}{\partial x_{i}} \,,\quad
\lim_{\tau\to0} \,\frac{\partial {\cal S}_{5}}{\partial y_{i}} \,,\quad
\lim_{\tau\to0} \,\frac{\partial {\cal S}_{5}}{\partial z_{i}} \,,\quad
\lim_{\tau\to0} \,\frac{\partial {\cal S}_{5}}{\partial w_{i}} \right\}
\right|_{\eqref{para4}}=0,  \quad \text{for ~} i=1,\ldots 8.
\ee
Among the above $32$ equations, only 8 of them are independent. One can find that there are $24$ solutions by solving the system above. We can also easily eliminate $u$, $v$, $p$ , $q$ and $r$ in the new scattering equations \eqref{neweq581} and then reduce the system to one only involving hard particles
\be\label{pino}
\left. \left\{\frac{\partial \tilde{\cal S}_{5}}{\partial y_{7}}+ \frac{x_7-z_7}{y_7-z_7} \frac{\partial \tilde{\cal S}_{5}}{\partial x_{7}} ,\quad
\frac{\partial \tilde{\cal S}_{5}}{\partial z_{7}}+ \frac{y_7-x_7}{y_7-z_7} \frac{\partial \tilde{\cal S}_{5}}{\partial x_{7}} ,\quad
\frac{\partial \tilde{\cal S}_{5}}{\partial w_{7}}+ \frac{\left(x_7-z_7\right){}^2}{z_7-y_7}\frac{\partial \tilde{\cal S}_{5}}{\partial x_{7}}
 \right\}\right|_{w_{7},y_{8},w_{8}\to \frac{y_{7}-z_{7}}{x_{7}-z_{7}}}=0,
\ee
where ${\tilde{\cal S}_{5}}$ is the potential for hard particles,
$
{\tilde{\cal S}_{5}}\equiv \sum_{1\leq a<b<c<d<e\leq 7}s_{abcde}\log |a, b , c,d,e|
$. In \eqref{pino} we presented only three independent equations for the three remaining variables $y_{7}$, $z_{7}$ and $w_{7}$. Notice that even though the equations in \eqref{pino} are different from the original scattering equations for $X(5,7)$, they share the same number of solutions.

It is not obvious that there are solutions for the new equations \eqref{neweq581}.  Let's use the positive kinematic data to clarify it.  Recall that, in the soft limit for $X(3,7)$,  the soft particle is bounded by a $2$-simplex, i.e. a triangle, formed by three lines in ${\mathbb {RP}}^{2}$. For $X(4,7)$, the soft particle is bounded by a 3-simplex, i.e. a tetrahedron, formed by four planes in ${\mathbb {RP}}^{3}$ space. 
It turns out that for $X(5,8)$, we can geometrically interpret the second configuration in table \ref{458} as having the soft particle bounded by a 4-simplex formed by five $3$-planes in ${\mathbb {RP}}^{4}$. Its four projections are shown in figure \ref{11fourf}.

\begin{figure}[!htb] 
\centering
\begin{tikzpicture} [xscale=1,yscale=1]

\draw (0.98,-4.67)node {.} node[left] {$\{9,10\}$}--
(3.44,-5.68) 
node {.}  node[below] {12}
--(3.60,-3.20)node {.}  node[right] {13}
--(2.19,-3.42)node {.}  node[above] {11}
--(0.98,-4.67)
--(3.60,-3.20)
(2.19,-3.42) --(3.44,-5.68)

(2.29,-4.25) --(2.29,-4.25) node{.}node[below]{8}
;

\draw (4.96,-5.41)node {.} node[left] {9}--
(6.48,-6.06) 
node {.}  node[below] {\{10,12\}}
--(7.58,-4.15)node {.}  node[right] {11}
--(6.93,-3.15)node {.}  node[above] {13}
-- (4.96,-5.41)
--(7.58,-4.15)
(6.93,-3.15) --(6.48,-6.06)

(6.58,-4.87) --(6.58,-4.87) node{.}node[below left]{8}
;

\draw (8.65,-4.90)node {.} node[left] {9}--
(9.67,-5.33) 
node {.}  node[below] {\{10,11\}}
--(10.74,-3.84)node {.}  node[right] {12}
--(10.91,-1.84)node {.}  node[above] {13}
-- (8.65,-4.90)
--(10.74,-3.84)
(10.91,-1.84) --(9.67,-5.33)

(10.05,-4.05) --(10.05,-4.05) node{.}node[ left]{8}
;

\draw (11.25,-5.46)node {.} node[left] {9}--
(12.76,-5.90) 
node {.}  node[below] {10}
--(13.60,-5.31)node {.}  node[right] {11}
--(13.43,-2.89)node {.}  node[above] {\{12,13\}}
--(11.25,-5.46)
--(13.60,-5.31)
(13.43,-2.89) --(12.76,-5.90)

(12.99,-4.18) --(12.99,-4.18) node{.}node[ left]{8}
;

\end{tikzpicture}

View from particle 2\quad View from particle 3 \quad View from particle 4\quad View from particle 5
\caption{Geometrical interpretation for the topology type 2 near the soft limit. The soft particle is bounded by a 4-simplex. Here we show four projections  of the 4-simplex from the viewpoint of particles 2, 3, 4 and 5, respectively. In the strict soft limit the 4-simplex collapses to a point where the soft particle lies.}
\label{11fourf}
\end{figure}

The five vertices of the 4-simplex can be seen as auxiliary points, each of them having a label ranging from 9 to 13. The five  facets of the 4-simplex, each of them corresponding to a  tetrahedron, have vertices labelled by $\{9, 10, 12, 13\},$ $ \{9, 10, 11, 13\}$, $\{9, 11, 12, 13\}$, $\{9, 10, 11, 12\}$ and $\{10, 11, 12, 13\}$, respectively. They are passed by the five dominating 3-planes  $\overline{1 2 3 5} $, $\overline{1 2 4 6}$, $ \overline{1 5 6 7}$, $\overline{234 7}$ and $\overline{3 4 5 6}$, respectively. 

In the first projection shown in figure \ref{11fourf}, two points $\{9,10\}$  are pinched from the viewpoint of particle 2, which has been sent to infinity.  We can say that particle 2 lies on the line determined by \{9,10\}. Particles 1 and 6 lie on the line determined by \{9,13\} and \{11,13\} respectively. Particle 7  lies on a 2-plane determined  by three vertices of the 4-simplex $\{9, 11, 12\}$. The remaining three projections are completely analogous.

The polytopes in
the remaining two configurations in table \ref{458} are slightly more complicated.  Actually, now there are two bounded chambers formed by the dominating 3-planes for the configurations type 3 and type 4. The soft particle can lie in either of the two bounded chambers.  See appendix \ref{appendix2} for more details. In the following, we just show how to get the solutions for general kinematic data.

For type 3, we make a reparameterization of \eqref{cp4} under the constraints
\be
|1 2 3 7 8|=\tau u,\,\, |1 2 4 5 8|=\tau v, \,\,|13568|=\tau p,\,\,|2346 8|=\tau q, \,\,|2  56 78|=\tau r,\,\,|345 78|=\tau s\,,
\ee
that is
\begin{align}\label{8solu}
&y_7= \frac{\tau(s +r  -  v+  v
   z_7-  p)+x_7-z_7}{ \tau (s-  p)+x_7-1},
   \quad
   w_7= \frac{ \tau(q
    z_7+  u)-z_7}{\tau(p-s )-x_7}, \nonumber
    \\
    &x_8=  \tau s
   +x_7, \quad y_8= \tau  v,\quad z_8=  \tau (s -  p)+x_7,\quad w_8= 1- \tau q.
\end{align}
When this is plugged into the original scattering equations and the strict soft limit $\tau\to 0$ is taken, we obtain a new set of scattering equations which has 8 solutions.

Likewise for type 4, we make a reparameterization of  \eqref{cp4} under the constraints
\be
|1 2 3 4 8|=\tau u,\,\, |1 2 3 5 8|=\tau v, \,\,|123 78|=\tau p,\,\,|1456 8|=\tau q, \,\,|2 4 5 78|=\tau r,\,\,|346 78|=\tau s\,,
\ee
that is
\begin{align}
&z_7= \frac{ \tau (r v-r   p+q   v-q
   p+  u v+  s v-  u v x_7)+v x_7-v y_7+w y_7-w}{u
   \left(- \tau (r+q) +y_7-1\right)}, \nonumber
   \\
   &w_7= \frac{\tau (r  +q
   +  u+  s-  u x_7)+x_7-y_7}{ \tau(r +q)
   -y_7+1}, \nonumber
   \\
   &x_8= -  \tau(q+r )  +y_7, \hspace{5mm} y_8= y_7- \tau r ,  \hspace{5mm}z_8=
   -\tau  v,  \hspace{5mm}w_8= \tau  u\,,
\end{align}
which will make $|1 2 3 6 8|=\tau (u+v)$ vanish as well. When we plug this into the original scattering equations and take the strict soft limit $\tau\to 0$, we obtain a new set of scattering equations which has 32 solutions.

As mentioned before, combining all these results we obtain ${\cal N}_8^{(5):{\rm  singular}}=420\times 24+ 210\times8+ 840\times32 + 210\times 96=58\, 800 $ singular solutions. In addition to the already known ${\cal N}_8^{(5):{\rm regular}}=24\times 5388=129\,312$ regular solutions, the total number of solutions for $X(5,8)$ is ${\cal N}_8^{(5)}=188\,112$, which is exactly the same as the result obtained for $X(3,8)$.


\section{General Configurations that Support Singular Solutions}\label{sec6}

In this section we explain what we believe are all configurations of points that lead to singular solutions in soft limits $X(k,n)\to X(k,n-1)$ for general $k$ and $n$. Our proposal is based on the examples already computed and on many other configurations for which we have been able to compute particular solutions.

In general, recall that singular solutions will make some minors involving the soft particle vanish while keeping all other minors non-vanishing. In particular, no subset of only hard particles should develop any linear dependence detected by the vanishing of a single minor as this would imply that the hard-particle kinematics is not generic. We start by assigning each hard particle a position in the ${\mathbb {CP}}^{k-1}$ and each vanishing minor will correspond to a $(k-2)$-plane determined by $(k-1)$ particles. 

We state our conjecture distinguishing $k=3$ from $k>3$. The reason is that $k=3$ is the base case for the rest since $k=2$ does not have singular solutions.

\begin{figure}[H]
\centering
\includegraphics[width=150mm]{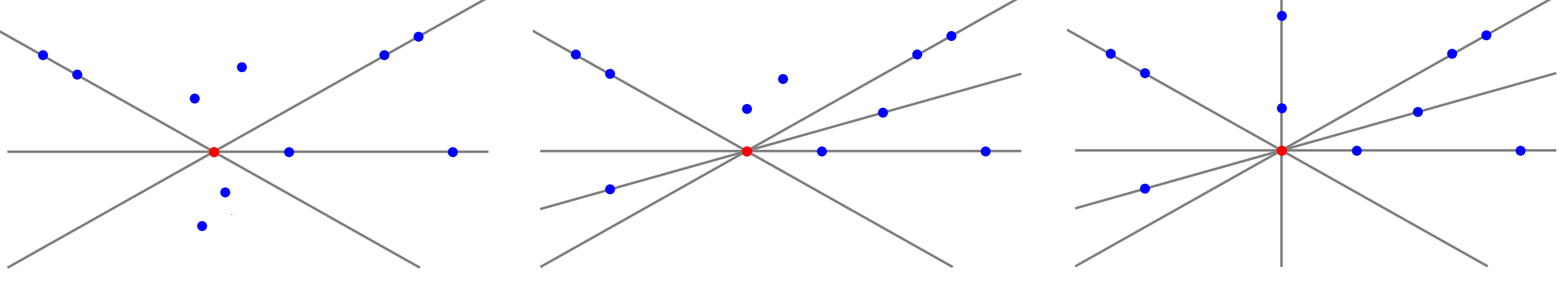}
\caption{All configurations of singular solutions in $n=11$. The soft particle is represented as a red point and the hard particles are represented as blue points. \textit{Left}: one possible situation is when we have 3 vanishing minors involving the soft particle. Namely, we have three lines, each one passing through two hard particles, intersecting at the soft particle. The rest of the hard particles do not develop any linear dependence. \textit{Center}: another possible situation is when we have 4 vanishing minors involving the soft particle, i.e. 4 lines. \textit{Right}: the last possibility in $n=11$ is when we have $m=5$ vanishing minors involving the soft particle, i.e. 5 lines.}
\label{n11}
\end{figure}

For $k=3$ and $n=2m+1$ or $n=2m+2$ with $m\geq 3$, we conjecture that singular solutions come from configurations where $3,4,\cdots, m$ lines meet at the soft particle respectively. Each line is determined by two hard particles and of course no subset of three hard particles are allowed to be collinear. We have checked that up to $n=14$ there are indeed solutions supported by all such configurations. In figure \ref{n11}, we show all three configurations for $n=11$.

For $k\geq 4$, we conjecture that singular solutions come from two kinds of configurations. The first kind of configurations is obtained from the cases with lower $k$ and $n$. More explicitly, this kind requires that all vanishing minors share a set of hard particles. Besides, the remaining hard particles together with the soft particle are projected by the common hard particles to a lower dimension space and their projections satisfy the requirement of that particular  dimension. 

For example, we already know there are singular solutions where three minors of the form $|14n|$, $|25n|$ and $|36n|$ vanish for $k=3$ and $n\geq7$. Thus we expect that for any $k\geq 4$ and  $n\geq k+4$, there will always be solutions coming from configurations  where three minors of the form $|1478\cdots k+3,n|$, $|2578\cdots k+3,n|$ and  $|3678\cdots k+3,n|$ vanish.
 The hard particles 1-6   together with the soft particle are projected to a lower dimension space through the hard particles $7$, $8$, $\cdots$, $k+3$ one by one.  Finally, they are  projected  to ${\mathbb {CP}}^{2}$ and  the projection of the soft particle is crossed by three lines. 
We have  numerically checked that for any $n\leq12$ and $4\leq k\leq n-4$, there are indeed solutions of this kind.

Similarly, we already know there are singular solutions where four minors of the form $|123n|$, $|345n|$, $|561n|$ and $|246n|$ vanish for $k=4$ and $n=7$. Thus we expect that for any $k\geq 4$ and $n\geq k+3$, there will always be solutions coming from configurations  where four minors of the form $|12378\cdots k+2,n|$, $|34578\cdots k+2,n|$,
 $|56178\cdots k+2,n|$ and  $|24678\cdots k+2,n|$ vanish.
 The hard particles 1-6   together with the soft particle are  projected to a lower dimension space through the hard particles $7$, $8$, $\cdots$, $k+2$ one by one. Finally, they are  projected to ${\mathbb {CP}}^{3}$ and the projection of the soft particle is  crossed by four planes. 
We have  numerically checked that for any $n\leq 11$ and $4\leq k\leq n-3$, there are indeed solutions of this kind.

The second kind of configurations correspond to those
where at least $k$ $(k-2)$-planes meet at the soft particle location.  
Each of the  $(k-2)$-planes are determined by $k-1$ hard particles.
Besides,  by slightly changing the position of hard particles, these $(k-2)$-planes can form a  polytope with infinitesimal volume  around the position of the soft particle without any hard particle as one of its vertices.  
 Of course, no subset of $k$ hard particles can lie on a single $(k-2)$-plane as this would imply an unwanted linear dependence. In particular, any configuration that supports singular solutions must still support singular ones for higher $n$ and the same $k$.

For example, for $k=4$ and $n=8$, we find 7 different topologies of configurations  that satisfy the requirements to support singular solutions, as shown in table \ref{4448}.
\begin{table}[H]
\begin{tabular}{llll}
 I& $|1478|,|2578|,|3678|$
 \\
II& $|1238|,|3458|,|5618|,|2468|$& \!\!\!\!\!\!\!\! III& $|1238|,|3458|,|5678|,|2468|
 $  
 \\
IV& $|1 4 78|, |25 78|, |3 6 78|, |1 238|, |4568|$ &\!\!\!\!\!\!\!\! V& $|1238|,|1458|,|1678|,|2468|,|2578|$  \\ VI&
 $|1238|,|1458|,|1678|,|2468|,|2578|,|3478|$ 
 \\ VII&
 $|1238|,|1248|,|1258|,|1268|,|1278|,|3458|,|3678|$
\end{tabular}
\caption{Seven different topologies of singular configurations for $X(4,8)$ defined by the list of vanishing minors.}
\label{4448}
\end{table}
We have numerically checked that there are solutions for each of them.  
  
The first topology belongs to the first kind of configurations which are related to that of $X(3,7)$ through projection.  For the remaining six topologies,  there are at least 4 planes meeting at the soft particle location. 

The second topology is the same configuration that supports the singular solutions in $X(4,7)$. The third topology has one hard particle changed in the third vanishing minor with respect to the second topology.  Note that  no matter what hard particle in a single minor in the second configuration is changed as 7, they all  lead to the configuration of the same topology.  

Comparing the first and  fourth topology, we see in addition to three common minors $|1478|,|2578|,|3678|$,  the fourth topology has two more vanishing ones $|1 238|, |4568|$, which supports singular solutions. However,  if we just add one more vanishing minor, such as  $|1238|$, to those of the first topology,  there will be no singular solutions. This is because by slightly changing the position of hard particles, the four planes $\overline{147}$, $\overline{257}$, $\overline{367}$ and $\overline{123}$ can form a  polytope with infinitesimal volume around the position of the soft particle but with the hard  particle 7 as one of the  vertices, which is forbidden. 

Another way to think about it is to use the geometry description using positive kinematic data \eqref{potential4}.  
For the first  configuration, three planes $\overline{147}$, $\overline{257}$ and $\overline{367}$ share a common line that crosses particles 7 and 8.  The final position of the soft particle 8 in the line is determined by both vanishing and finite minors.  However, we can also fix the position of 8 in the line by imposing two more vanishing minors. That is the case of the fourth configuration.  One can imagine that we cannot just impose one more vanishing minor as this will push the particle 8 in the line into infinity, i.e. there will be no singular solutions where only 4 minors of the form $|1478|$, $|2578|$, $|3678|$ and $|1238|$ vanish.  

We have also checked many other examples up to $k=6$, all of them having solutions (see table \ref{458fajio}).

\begin{table}[!htb]
\centering
\begin{tabular}{c|c|l}
 $k$&$n$&Vanishing minors  \\
 \hline
 4&9& $ |1 2 3 n | , |3 4 5 n | , |5  6 7  n  | , |7 8 1 n|$  \\
  4&9&
 $|1  2  3  n|, |1  2  4  n|, |1  2  6  n|, |1  2  7  n|, |3  4  5  n|, |3  6  8  n|$\\
4&10& $ |1 2 3 n | , |3 4 5 n | , |5  6 7  n  | , |7 8 9 n|$  \\
4&10& $
|1  2  3 n|, |1  4  5 n|, |1  6  7 n|, |2  4  6 n|, |2  5  7 n|, |3  4  7 n|, |3  8  9 n|$

  \\
5&9& 
  
$ | 1  2  3  5  n  | , | 1  2 4  6  n  | , | 1 5 6  7 n  | , | 2  3  4  7 n  | , | 3  4  5  8 n |
  $
  
    \\
6&10& 

 $ | 1  2  3  5  8  n | , | 1  2  4  6  8  n | , | 1 5 6  7 8  n  | , | 2  3  4  7 8  n  | , | 3  4  5  8 9  n |
  $
\end{tabular}
\caption{Some explicit examples supporting singular solutions, where particle $n$ is the soft one.}
\label{458fajio}
\end{table}

\section{Discussions} \label{sec7}

In this work we have started the study of singular solutions in soft limits. This is a new phenomenon for scattering equations on $X(k,n)$ with $k>2$. We computed all singular solutions for all cases with $n<9$, except for $X(4,8)$. This proved that studying singular solutions is an effective technique for computing the number of solutions in cases where other known techniques cannot be applied. For example, we have proven that ${\cal N}_8^{(3)}={\cal N}_8^{(5)}=188\,112$. Also, even in cases where indirect approaches are possible, singular solutions prove to be a much simpler route as seen in the alternative determination of ${\cal N}_7^{(3)}={\cal N}_7^{(4)}=1\,272$. 

One of the most pressing issues is to extend our study to all $X(3,n)\to X(3,n-1)$ cases with $n>8$. In section \ref{sec6}, we presented a conjecture for all configurations that can support singular solutions. It is very tempting to suggest that in this case, it would be possible to count solutions using a recursive approach. Recall that in $X(3,8)\to X(3,7)$ we resorted to a second soft limit in order to count solutions. Such ``fibration'' structure is familiar in the $k=2$ case where $X(2,n)$ can be thought of as a fibration over $X(2,n-1)$ (see e.g. \cite{Mizera:2019gea}). For $k=3$ the structure we have uncovered is much more interesting and we leave its study for future work.      

The scattering equations have been a powerful tool for studying properties of scattering amplitudes via the CHY formalism \cite{Cachazo:2013hca,Cachazo:2013iea}. The quantum field theory whose amplitudes have the simplest CHY formulation is a theory with a $U(N)\times U(\tilde N)$ flavour group and a scalar field in the biadjoint representation with cubic interactions (for related developments see e.g. \cite{Cachazo:2014hca,Broedel:2013,Dolan:2014,Mizera:2017inverse,Mizera:2017KLT,ABHY,Frost:2018,Salvatori:2018,BLR:2019,Raman:2019,Drummond:2019qjk,Borges:2019}). It is not surprising that this is the first theory to have been generalized so that it has a CHY representation based on $X(k,n)$ with $k>2$ \cite{Cachazo:2019apa}. We now turn to a discussion on such biadjoint amplitudes and their soft limit behavior on singular solutions using the explicit cases we have computed and the conjecture regarding the general configurations that can support them.     

\subsection{Generalized Biadjoint Scalar Soft Limit}

Recall the generalized biadjoint scalar amplitude \cite{Cachazo:2019apa}

\be
m^{(k)}_{n}[\alpha|\beta]=\int \Bigg[\frac{1}{{\text{Vol}}[\mathrm{SL}(k, \mathbb{C})]}\prod_{a=1}^n\prod_{i=1}^{k-1}dx_a^i\Bigg]\prod_{a=1}^n\sideset{}{'}\prod_{i=1}^{k-1}\delta\left(\frac{\partial {\cal S}}{\partial x_a^i}\right)\text{PT}^{(k)}_n[\alpha]\text{PT}^{(k)}_n[\beta],
\label{chy}
\ee
where the Parke-Taylor functions correspond to

\be
\text{PT}^{(k)}_n[12\cdots n]=\frac{1}{|12\cdots k||23\cdots k+1|\cdots|n1\cdots k-1|}.
\label{PT}
\ee
Now consider the soft limit for one particle, for instance $s_{abn}=\tau\hat{s}_{abn}$.

Following our conjecture in section \ref{sec6}, one can analytically show that when $\tau\to0$ the singular solutions for $k=3$ and general $n$ can at most contribute to order ${\cal O}(\tau^{-1})$ to the amplitude. The argument goes as follows. For $k=3$ and $n\geq7$ we have seen that there is always one singular configuration with $3$ vanishing minors involving the soft particle. If we choose e.g. particle $n$ to be the soft one, we parameterize the vanishing minors as $|abn|\sim u\tau$. This means that the Jacobian for the change of variables will give an ${\cal O} (\tau^3)$ factor in the amplitude\footnote{Note that for singular configurations with $m$ vanishing minors the Jacobian gives orders of ${\cal O} (\tau^{m})$. That's why having only 3 vanishing minors corresponds to the leading contribution.}. Moreover, given the form of the singular configurations, we can at most have two vanishing determinants in the Parke-Taylor functions. This produces an ${\cal O} (\tau^{-4})$ factor in the amplitude. Hence, the leading contribution of the singular solutions to the biadjoint scalar amplitude in $k=3$ is of order ${\cal O} (\tau^{-1})$.

Now we move on to explain the contribution to the amplitude for the cases $k=4$ and $n=7$, and $k=5$ and $n=8$ in the soft limit. Consider again the biadjoint scalar amplitude (\ref{chy}). For $k=4$ and $n=7$ the singular configurations are those where 4 vanishing minors involve the soft particle. The Jacobian for the change of variables thus gives an order ${\cal O} (\tau^{4})$ to the amplitude. From the Parke-Taylor functions (\ref{PT}) we again obtain a factor of ${\cal O} (\tau^{-4})$,  hence the contribution to the amplitude in this case is of the order ${\cal O} (\tau^{0})$.

For $k=5$ and $n=8$ the analysis is slightly different. 
In this case the configuration that gives a more dominant contribution to the amplitude is the one with only 4 vanishing minors. Following the same procedure as for $k=4$ and $n=7$, this would naively give us again a total contribution of order ${\cal O} (\tau^{0})$ to the amplitude. However, recall that the leading order in $\tau$ for one of the soft scattering equations vanished in this configuration. This means that we get an extra factor of ${\cal O} (\tau^{-1})$ in the amplitude, coming from the subleading term of the vanishing soft scattering equation \eqref{sub}. Therefore, the contribution to the amplitude in this case is of the order ${\cal O} (\tau^{-1})$\footnote{ Singular configurations with 5, 6 and 7 vanishing minors give  orders of ${\cal O} (\tau^{1})$, ${\cal O} (\tau^{2})$ and ${\cal O} (\tau^{1})$, respectively.}.

We haven't obtained the whole set of singular solutions for higher values of $k$ and $n$, but in what follows we make a prediction on their contribution to the biadjoint scalar amplitudes in the soft limit expansion based on our conjecture in section \ref{sec6}.


For any $k\geq 4$ and $n\geq k+3$, as we have already explained, there will always be some singular solutions from configurations where four minors of the form $|12378\cdots k+2,n|$, $|34578\cdots k+2,n|$,
 $|56178\cdots k+2,n|$ and  $|24678\cdots k+2,n|$ vanish. We can use the gauge redundancy of  ${\mathrm{SL}}(k,{\mathbb C})$   to send $7$, $8$, $\cdots$, $k+2$ to infinity in different directions.  A direct consequence is that the leading order for the scattering equations of the soft particle corresponding to these directions vanish.  See section \ref{vanishleading} as an example.   This means that we get a factor of ${\cal O} (\tau^{4-k})$  from the subleading term of the vanishing soft scattering equations.  The Jacobian for the change of variables in this case gives an order ${\cal O} (\tau^{4})$ and the Parke-Taylor functions (\ref{PT}) also give a factor of ${\cal O} (\tau^{-4})$. Thus we expect that  the contribution to the amplitude in this case is at most of the order ${\cal O} (\tau^{4-k})$. Besides $X(5,8)$, we have numerically checked the existence of such kind of solutions for $X(6,9)$.
 
 Similarly, for any $k\geq 3$ and $n\geq k+4$, singular solutions from configurations where three minors of the form $|1278\cdots k+3,n|$,
 $|3478\cdots k+3,n|$ and  $|5678\cdots k+3,n|$ vanish will contribute to the amplitude  with order ${\cal O} (\tau^{2-k})$  at most. The reason is as follows. Consider the second kind of configurations conjectured in section \ref{sec6}.  The Jacobian for the change of variables in this case gives an order ${\cal O} (\tau^{k})$ 
at most since there are at least $k$ vanishing minors.  
Although any 
Parke-Taylor function has $k$ minors involving the soft particles, there can be at most $(k-1)$ vanishing ones, otherwise some minors only involving hard particles must vanish.
Since in this case all the scattering equations keep their leading terms, the upper bound of the contribution of the singular solutions is  of order ${\cal O} (\tau^{2-k})$.
 We have numerically checked the existence of such kind of solutions for $X(4,8)$ and $X(5,9)$.



The leading contribution to the biadjoint scalar theory amplitude is ${\cal O}(\tau^{1-k})$. This actually implies that singular solutions do not contribute to leading order. We summarize these results in table below\footnote{We thank A. Guevara for results and comments on this table.}:

\begin{table}[!htb]  
  \centering
  \begin{tabular}{ |p{5cm}|p{2.5cm}|p{2.5cm}|p{2.5cm}|}
    \cline{2-4}
    \multicolumn{1}{c|}{} &\hspace{7mm}\textbf{k = 3} &\hspace{7mm}\textbf{k = 4}
    &\hspace{7mm}\textbf{k = 5}\\
 \hline
 Regular solutions   & \hspace{7mm}${\cal O}(\tau^{-2})$& \hspace{7mm}${\cal O}(\tau^{-3})$& \hspace{7mm}${\cal O}(\tau^{-4})$\\
  \hline
Singular solutions for \textbf{n = 7} & \hspace{7mm}${\cal O}(\tau^{-1})$& \hspace{7mm}${\cal O}(\tau^{0})$
& \hspace{11mm} -\\
\hline
Singular solutions for \textbf{n = 8} & \hspace{7mm}${\cal O}(\tau^{-1})$ &
\hspace{7mm}{\color{red}${\cal O}(\tau^{-2})$}& \hspace{7mm}${\cal O}(\tau^{-1})$\\
\hline
Singular solutions for  \textbf{n $\geq$ 9} & \hspace{7mm}{\color{red}${\cal O}(\tau^{-1})$} &
\hspace{7mm}{\color{red}${\cal O}(\tau^{-2})$}& \hspace{7mm}{\color{red}${\cal O}(\tau^{-3})$}\\
\hline
  \end{tabular}
  \caption{Leading order contribution in the soft limit expansion. The results in black are got from analytic derivation. The results in {\color{red}red} come from what we conjecture.}
  \label{softorder}
  \end{table}

From table \ref{softorder} one can notice an interesting pattern. For $k\geq4$ the singular solutions for $n=k+3$ do not contribute to the subleading term. For $n>k+3$, though, the singular solutions will always be relevant, i.e. will contribute to the subleading term in the biadjoint scalar amplitude. This special case $k\geq4$ and $n=k+3$ is nothing but the case that just comes after $n=k+2$, i.e. when no singular solutions arise. This phenomenon does not appear in $k=3$ since for $n=6$ there are no singular solutions, as explained before. Therefore, for $k=3$ the singular solutions will always contribute to the subleading term.

These results in fact resonate with the recent work of Garcia and Guevara \cite{GG}. More precisely, they computed the leading order behavior of biadjoint scalar amplitudes in the limit when a soft particle decouples from the scattering equations of the hard particles. Hence, no singular configurations were taken into account. With this assumption, they found that the leading soft factor for the $m^{(k)}_n$ amplitude is 

\be
S^{(k)}_n=m^{(k)}_{k+2}(\mathbb{I}|\mathbb{I})(\hat{\textbf{T}}^{(p,q)}_{k+2}\to\hat{\textbf{T}}^{(p,q)}_{n})
\label{ggsoft}
\ee
where the canonical ordering is assumed and
\be
\hat{\textbf{T}}^{(p,q)}_{n}:=\sum_{a_1,...,a_r=1}^{n}\hat{\textbf{s}}_{12...qa_1...a_r(n-k+q+r+1)...n-1n}
\ee
are planar kinematic invariants, $0\leq r\leq k-2$ and $1\leq q\leq k-r$, and $r$ denotes the number of summed indices. The fact that singular solutions do not contribute to the leading order in the soft limit expansion of the biadjoint scalar amplitudes serves to corroborate their statement \eqref{ggsoft} for the cases already mentioned. Indeed, they numerically checked that only regular solutions contribute to leading order for $k=3$ and $n=7$, $8$, for $k=4$ and $n=7$ and for $k=5$ and $n=8.$

\section*{Acknowledgements}

We would like to thank A. Guevara and J. Rojas for very useful discussions and especially S. He and T. Lam for sharing their proposal on the relation between the number of representations of uniform matroids and the number of solutions to scattering equations.  Research at Perimeter Institute is supported in part by the Government of Canada through the Department of Innovation, Science and Economic Development Canada and by the Province of Ontario through the Ministry of Economic Development, Job Creation and Trade.

\renewcommand{\thefigure}{\thesection.\arabic{figure}}
\renewcommand{\thetable}{\thesection.\arabic{table}}

\appendix

\section{Singular Solutions in $X(3,n)$ from Bounded Chambers Counting\label{appendix1}}

In this appendix we show how to visualize and count the number of singular solutions in $X(3,7)$ and the number of $({\texttt {regular}}_7,\texttt{singular}_8)$ solutions in X(3,8) with positive kinematics.

\subsection{Singular Solutions in $X(3,7)$\label{appendix12}}

We have seen in section \ref{x37} that with positive kinematic data all the solutions we obtain are real. This means we can analyze them by counting bounded chambers in $\mathbb{R}\mathbb{P}^2$ space when $|147|$, $|257|$ and $|367|$ vanish. We expect to find 12 bounded chambers, which would correspond to the 12 solutions for each of the 15 existing configurations. 

\begin{figure}[!htb]
\centering
\includegraphics[width=150mm]{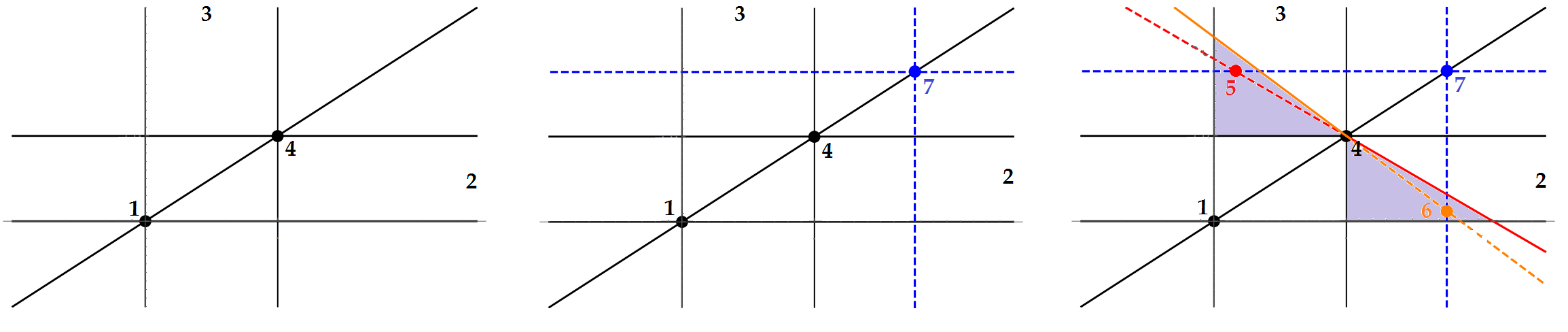}
\caption{\textit{Left}: the first four particles are gauge-fixed. This creates 5 repelling lines, drawn in black, and particle 7 must be on the line that passes through 1 and 4. \textit{Center}: we now consider the situation in which the soft particle 7 is in the outside-right(left) of the square $[0,1]^2$. \textit{Right}: particles 5 and 6 must lie on the blue dashed lines created by particles 7, 2 and 3. This only happens if both particles bound each other through particle 4(1) (red and orange lines). The two grey bounded chambers are those where particles 5 and 6 can be.}
\label{s1}
\end{figure}

The bounded chambers come in the following way. First, we use the same gauge fixing for the first four particles as explained in section \ref{x37}. This creates 5 repelling lines, one of them crossing the diagonal of the square $[0,1]^2$ created by particles 1 and 4. It is precisely on this line where the soft particle 7 must be. We can have solutions where particle 7 is outside the square $[0,1]^2$, since particles 5 and 6 can simultaneously create bounded chambers for each other. We represent this situation in figure \ref{s1}.

This configuration gives rise to 2 different solutions, since particles 5 and 6 can bound each other through particles 1 and 4 when particle 7 is outside the square $[0,1]^2$. 

\begin{figure}[H]
\centering
\includegraphics[width=150mm]{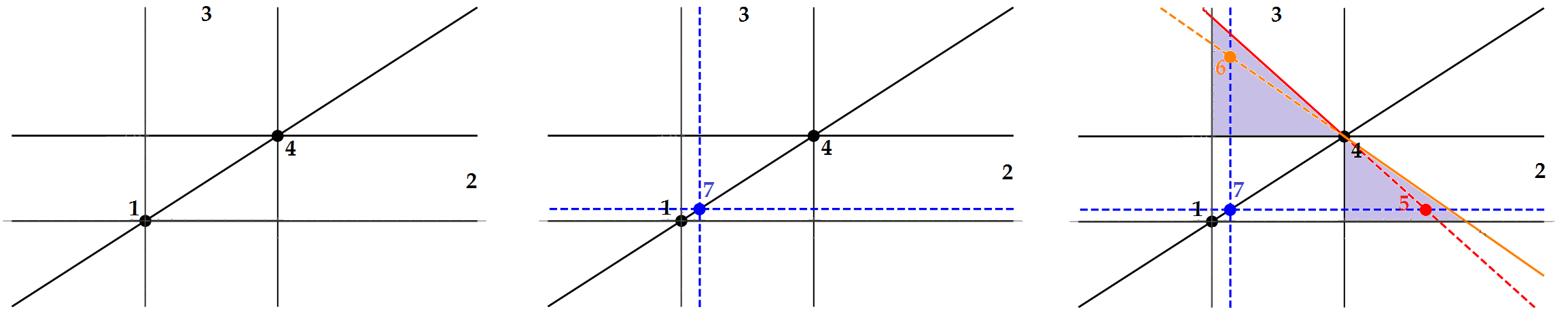}
\caption{\textit{Left}: the first four particles are gauge-fixed. This creates 5 repelling lines, drawn in black, and particle 7 must be on the line that passes through 1 and 4. \textit{Center}: we now consider the situation in which the soft particle 7 is inside the square $[0,1]^2$. \textit{Right}: particles 5 and 6 must lie on the blue dashed lines created by particles 7, 2 and 3. This only happens if both particles bound each other through particle 4(1) (red and orange lines). The two grey bounded chambers are those where particles 5 and 6 can be.}
\label{s2}
\end{figure}

Next, we also find solutions in the particular situation in which the soft particle 7 is inside the square $[0,1]^2$, but particles 5 and 6 are both outside of it. In this case, particles 5 and 6 also bound each other. We represent this situation in figure \ref{s2}.

This configuration also gives rise to 2 different solutions, since particles 5 and 6 can bound each other through particles 1 and 4. 

Finally, we also find solutions coming from having the soft particle 7 and the two remaining hard particles inside the square $[0,1]^2$. We represent this situation in figure \ref{ss3}:

\begin{figure}[!htb]
\centering
\includegraphics[width=150mm]{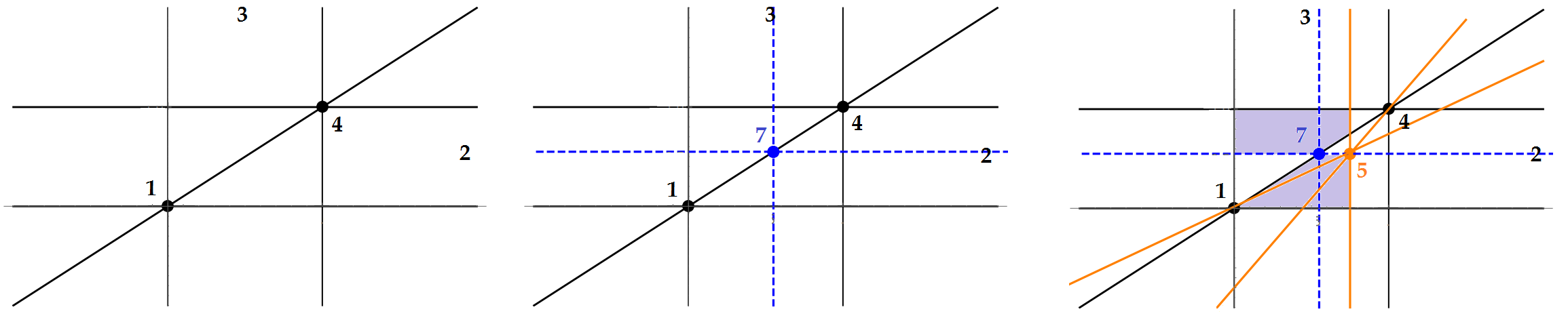}
\caption{\textit{Left}: the first four particles are gauge-fixed. This creates 5 repelling lines, drawn in black, and particle 7 must be on the line that passes through 1 and 4. \textit{Center}: we now consider the situation in which the soft particle 7 is inside the square $[0,1]^2$. This means that e.g. particle 5 must be in one of the two existing bounded chambers. \textit{Right}: particles 5 and 6 must lie on the blue dashed lines created by particles 7, 2 and 3. If we choose particle 5 to be e.g. in the lower-right bounded chamber, this creates 3 additional repelling lines, drawn in orange, which leave four bounded chambers where particle 6 can be, shown in grey.}
\label{ss3}
\end{figure}
This last situation gives rise to 3 solutions where both hard particles are in the same original bounded chamber, and 1 solution where both are in the different two original bounded chambers. Hence, there are a total of $2\times(3+1)=8$ solutions, since we can also choose particle 5 to be in the upper-left bounded chamber at first. Therefore, for this configuration, we count $2+2+8=12$ different solutions which correspond to the singular solutions already found before.

\subsection{$(\texttt{regular}_7,\texttt{singular}_8)$ Solutions in X(3,8)\label{appendix22}}

It turns out that the $(\texttt{regular}_7,\texttt{singular}_8)$ solutions studied in section \ref{sec51} are all real too. This opens the possibility to count them in $\mathbb{R}\mathbb{P}^2$ space in the same way as in appendix \ref{appendix12}. If we use the same gauge-fixing as in section \ref{x37} and consider the singular situation in which e.g. $|148|$, $|258|$ and $|368|$ vanish, we find ourselves in a similar fashion as in \ref{appendix12}, i.e. with 12 different situations. Yet, now we deal with one more particle (in this case particle 7) which is decoupled from the other hard particles. This particle can be found in 41 different equilibrium points, which gives the $12\times41=492$ solutions. Below we give an explicit visualization of one of the 12 different situations we can have:

\begin{figure}[!htb]
\centering
\includegraphics[width=110mm]{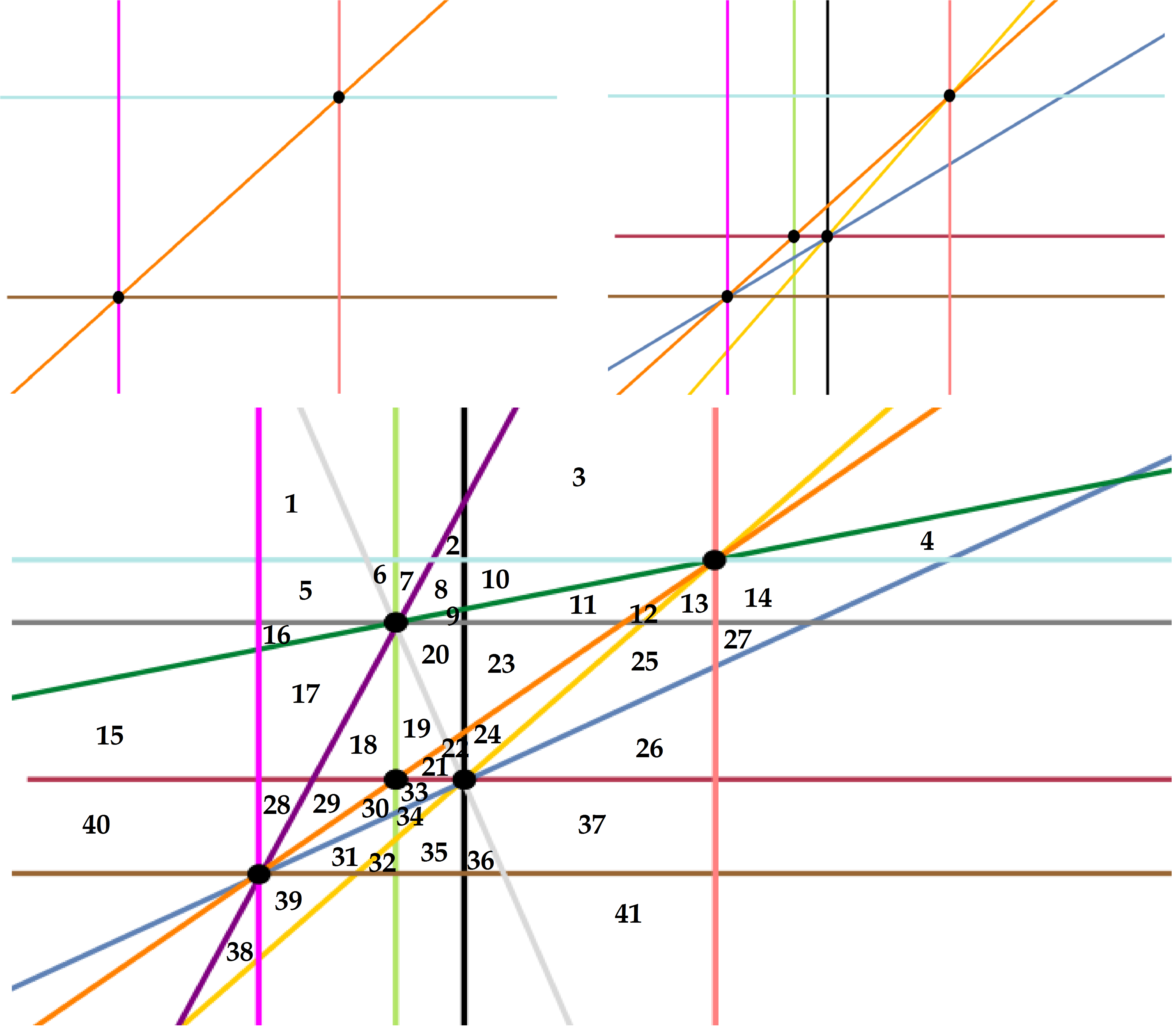}
\label{s3}
\caption{\textit{Top-Left}: particles 1, 2, 3 and 4 are gauge-fixed. This creates 5 repelling lines, and particle 8 must be on the line that passes through the two black points, which correspond to particles 1 and 4. Particles 2 and 3 are sent to infinity. \textit{Top-Right}: we now consider e.g. the third situation seen in \ref{appendix12}. The two new black points correspond to particles 5 and 8, and new repelling lines appear due to their interaction with the other particles. \textit{Bottom}: if we choose particles 5 and 6 to be e.g. on the two different original bounded chambers (see \textit{Top-Left} figure), this leaves us with 41 bounded chambers where particle 7 can be.}
\end{figure}

\section{Geometry Descriptions of Type 3 and 4 Configurations in $X(5,8)$ \label{appendix2}}

We can use the positive kinematic data to help us visualize the geometry underlying the singular solutions of the topologies type 3 and type 4 in table \ref{458}. 

For the topology type 3,  there are two bounded chambers formed by the six dominating 3-planes. See their projections in figure \ref{11fourf3}.  
\begin{figure}[!htb]
\centering
\begin{tikzpicture} [xscale=1,yscale=1]

\begin{scope}[xshift=0cm,yshift=0cm]
\draw[blue,thick] (0.99,-5.90)node {.} node[below] {\color{black}\{13,16\}}--
(2.91,-5.10) node {.}  node[below=3pt] {\color{black}\{10,11\}}
(1.26,-4.69)--(0.99,-5.05) node {.}  node[left] {\color{black}\{14,15\}}--(0.99,-5.90)
(0.99,-5.05) --(2.91,-5.10)
(0.99,-5.90)--(1.26,-4.69)
 ;

\draw[red,thick] 
(2.91,-5.10)- -(4.26,-4.51) 
node {.}  node[above] {\color{black}19}
--(4.12,-5.12)node {.}  node[below] {\color{black}20}--(2.91,-5.10)
(4.26,-4.51) --(1.43,-3.85) node {.}  node[above] {\color{black}18}
 --(1.42,-4.47) node {.}  node[above right] {\color{black}17}--(4.12,-5.12)
 (1.42,-4.47)--(1.26,-4.69) node {.}  node[left] {\color{black}\{9,12\}}--(1.43,-3.85)
;

\draw[green,thick] (1.26,-4.69)--(2.91,-5.10);

\draw[thick] (1.44,-5.18)--(1.44,-5.18)node {.}  node[below] {\color{black}8};
\end{scope}

\draw[blue,thick] (4.96,-7.31)node {.} node[below] {\color{black}13}--
(7.07,-7.16) node {.}  node[below] {\color{black}16}
--(7.37,-5.20)node {.}  node[right] {\color{black}15}--(5.15,-5.24) node {.}  node[left] {\color{black}14}--(4.96,-7.31)--(6.08,-4.71) node {.}  node[left=4pt] {\color{black}\{9,10\}}
--(5.15,-5.24) 
(7.37,-5.20)--(6.87,-4.72) node {.}  node[right=1pt] {\color{black}\{11,12\}}
--(7.07,-7.16)
 ;

\draw[red,thick] 
(6.08,-4.71)- -(6.75,-3.17) 
node {.}  node[above] {\color{black}\{18,19\}}
--(6.87,-4.72)
--(6.58,-4.44)node {.}  node[above right=4pt] {\color{black}\{17,20\}}--(6.08,-4.71)
(6.58,-4.44)--(6.75,-3.17) ;

\draw[green,thick] (6.08,-4.71)--(6.87,-4.72) ;

\draw[thick] (6.20,-5.49)--(6.20,-5.49)node {.}  node[below] {\color{black}8};

\begin{scope}[xshift=-.5cm]
\draw[blue,thick] (10.49,-6.06)node {.} node[below] {\color{black}9}
--(9.54,-5.87)node {.} node[below =5pt ] {\color{black}14}
--(8.85,-7.3)node {.} node[below] {\color{black}13}--
(10.36,-7.52) node {.}  node[below] {\color{black}16}
--(11.08,-6.17)node {.}  node[right] {\color{black}\{12,15,17\}}
(8.85,-7.3)--(9.61,-4.69)node {.} node[left] {\color{black}10}
--(9.54,-5.87)
(10.36,-7.52)--(10.14,-4.80)node {.} node[right] {\color{black}11}
 ;

\draw[red,thick] 
(10.49,-6.06)- -(11.59,-5.20) 
node {.}  node[above=3pt] {\color{black}18}
--(11.08,-6.17)
(11.59,-5.20) 
--(10.00,-3.27)node {.}  node[above ] {\color{black}19}--(9.63,-4.08)
--(9.61,-4.69)
(9.63,-4.08)node {.}  node[left ] {\color{black}20} --(10.14,-4.80)
(9.61,-4.69)--(10.00,-3.27)
--(10.14,-4.80) ;

\draw[green,thick] (9.61,-4.69)--(10.14,-4.80)--(11.08,-6.17)--(10.49,-6.06)--(9.61,-4.69) ;

\draw[thick] (10.05,-6.07)--(10.05,-6.07)node {.}  node[below] {\color{black}8};
\end{scope}
\begin{scope}[xshift=.2cm,yshift=-.5cm]
\draw[blue,thick](13.37,-4.80)node {.} node[left] {\color{black}10}-- (12.16,-6.42)node {.} node[below] {\color{black}\{13,14\}}--
(13.57,-6.54) node {.}  node[below] {\color{black}\{15,16\}}
--(13.88,-4.83)node {.}  node[right] {\color{black}11}
(12.16,-6.42)--(12.98,-5.71) node {.}  node[left=6pt] {\color{black}9}
(13.57,-6.54)--(13.57,-5.76) node {.}  node[right] {\color{black}12}  ;

\draw[red,thick] 
(12.98,-5.71)- -(13.57,-5.21) 
node {.}  node[right=4pt] {\color{black}\{17,18\}}
--(13.57,-5.76)
(13.57,-5.21)
--(14.04,-3.92)node {.}  node[above ] {\color{black}\{19,20\}}--(13.37,-4.80)
(14.04,-3.92)--(13.88,-4.83) ;

\draw[green,thick] (12.98,-5.71)--(13.57,-5.76)-- (13.88,-4.83)--(13.37,-4.80)--(12.98,-5.71);

\draw[thick] (13.04,-5.93)--(13.04,-5.93)node {.}  node[below] {\color{black}8};
\end{scope}

\end{tikzpicture}

View from particle 2\quad View from particle 3 \quad View from particle 4\quad View from particle 5
\caption{Four projections from the viewpoint of particles 2, 3, 4 and 5, respectively, of the two bounded chambers (shown in blue and red) for the topology type 3 in table \ref{458} and near the soft limit. Here we represent the case in which the soft particle is bounded by the blue chamber. The green edges correspond to shared edges by the blue and red chambers. In the strict soft limit, the two bounded chambers collapse to a point where the soft particle lies.
  \label{11fourf3} }
\end{figure}
The F-vectors of bounded chambers are both $\{8,16,14,6\}$.  
The 8 vertices of each bounded chamber are labelled by $\{9,10,11,12,13,14,15,16\}$ and $\{9,10,11,12,17,18,19,20\}$, respectively. For convenience, let's call the two bounded chambers as blue and red.

 Among the six facets of each bounded chamber,   two are  tetrahedrons and the remaining four  are truncated triangular prisms. The two  bounded  chambers don't share any facet but a dim-2 boundary of vertices $\{9,10,11,12\}$. Any dominating 3-plane passes through both facets of different bounded chambers, see table \ref{f3}.
\begin{table}[!htb]
\qquad\qquad\quad
\begin{tabular}{c|c|c}
${\rm Particles~ to~determine \atop dominating~3-planes}$ &${\rm Vertices ~of~ the ~facet~passed \atop by~the~ blue~chamber}$ &${\rm Vertices ~of~ the ~facet~passed \atop by~the~ red~chamber}$ \\
\hline
 \{1,2,3,7\} & \{9,10,11,12,13,16\} & \{9,10,11,12,18,19\} \\
 \{1,2,4,5\} & \{9,12,13,14,15,16\} & \{9,12,17,18\} \\
 \{1,3,5,6\} & \{9,10,13,14\} & \{9,10,17,18,19,20\} \\
 \{2,3,4,6\} & \{9,10,11,12,14,15\} & \{9,10,11,12,17,20\} \\
 \{2,5,6,7\} & \{10,11,13,14,15,16\} & \{10,11,19,20\} \\
 \{3,4,5,7\} & \{11,12,15,16\} & \{11,12,17,18,19,20\} \\
\end{tabular}
\caption{Dominating 3-planes and the facets they pass by in figure \ref{11fourf3}.\label{f3}}
\end{table}
Particles 1, 4, 6 and 7 lie in the lines that pass through \{9, 13, 18\}, \{12, 15, 17\}, \{10, 14, 20\} and  \{11, 16, 19\}, respectively.  Whilst particles 2, 3, and 5, which are sent to infinity, can be thought of as the intersections of four lines determined by four pairs of vertices. See   the first, second and fourth projections in figure \ref{11fourf3}.

The auxiliary points have proven to be very useful to understand the relative positions of the  hard particles. Alternatively, now we can ignore them and imagine how these hard particles form some dominating planes to bound the soft particle.

In the strict soft limit, the two bounded chambers collapse to a point. Some sets of four dominating 3-planes share a point where the soft particle lies, while some share a line. For example,  the four dominating 3-planes ${\overline {1237}}$,  ${\overline {1245}}$, ${\overline {2346}}$, and  ${\overline {2567}}$ share a common line where particles 2 and 8 lie.

There are 8  solutions of variables $u$, $v$, $p$, $q$, $r$, $s$, $x_{7}$ and $z_{7}$ for the new set of scattering equations, 
\be\label{neweq5811111}
\left. \left\{\lim_{\tau\to0} \, \frac{\partial {\cal S}_{5}}{\partial x_{i}} \,,\quad
\lim_{\tau\to0} \,\frac{\partial {\cal S}_{5}}{\partial y_{i}} \,,\quad
\lim_{\tau\to0} \,\frac{\partial {\cal S}_{5}}{\partial z_{i}} \,,\quad
\lim_{\tau\to0} \,\frac{\partial {\cal S}_{5}}{\partial w_{i}} \right\}
\right|_{\eqref{8solu}}=0,  \quad \text{for ~} i=1,\ldots 8.
\ee
These 8 solutions can be divided into four pairs. Although the two solutions of each pair are different, using the reparameterization \eqref{8solu},  they produce the same set of values for $\{x_7,y_7,z_7,$ $w_7,x_8,$ $y_8,z_8,w_8\}$, which corresponds to the fact that the two bounded chambers collapse to a single point.

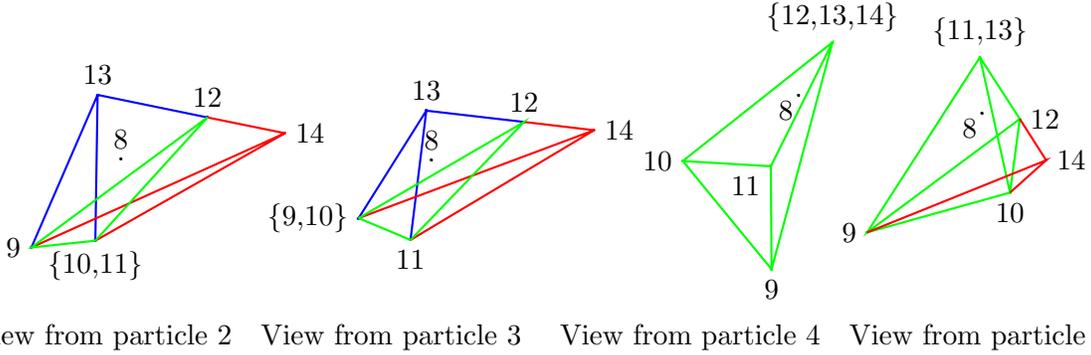
\begin{figure}[!htb]
\centering
 \begin{tikzpicture} [xscale=1,yscale=1]
 
\begin{scope}[xshift=0cm,yshift=0cm]
\draw[blue,thick](0.61,-4.06)node {.} node[left] {\color{black}9}-- (1.49,-2.03)node {.} node[above] {\color{black}13}--
(2.95,-2.33) node {.}  node[above] {\color{black}12}
(1.49,-2.03)
--(1.46,-3.97)node {.}  node[below] {\color{black}\{10,11\}}
 ;

\draw[red,thick] 
(0.61,-4.06)- -(3.98,-2.54) 
node {.}  node[right] {\color{black}14}
--(1.46,-3.97)
(3.98,-2.54) --
(2.95,-2.33) 
 ;

\draw[green,thick](1.46,-3.97)--(2.95,-2.33)-- (0.61,-4.06)--(1.46,-3.97);

\draw[thick] (1.80,-2.88)--(1.80,-2.88)node {.}  node[above] {\color{black}8};
\end{scope}

\begin{scope}[xshift=0.5cm,yshift=0cm]
\draw[blue,thick](4.46,-3.67)node {.} node[left] {\color{black}\{9,10\}}-- (5.36,-2.24)node {.} node[above] {\color{black}13}--
(6.66,-2.39) node {.}  node[above] {\color{black}12}
(5.36,-2.24)
--(5.15,-3.96)node {.}  node[below] {\color{black}11}
 ;

\draw[red,thick] 
(5.15,-3.96)- -(7.59,-2.50) 
node {.}  node[right] {\color{black}14}
--(4.46,-3.67)
(7.59,-2.50)--
(6.66,-2.39)
;

\draw[green,thick](4.46,-3.67)--(6.66,-2.39)-- (5.15,-3.96)--(4.46,-3.67);

\draw[thick] (5.43,-2.90)--(5.43,-2.90)node {.}  node[above] {\color{black}8};
\end{scope}

\begin{scope}[xshift=1.4cm,yshift=.6cm]

\draw[green,thick](7.87,-3.51)node {.}  node[left] {\color{black}10}--(9.05,-4.95)node {.}  node[below] {\color{black}9}-- (9.86,-1.93)node {.}  node[above] {\color{black}\{12,13,14\}}--(7.87,-3.51)
(7.87,-3.51)--(9.04,-3.58)node {.}  node[below left] {\color{black}11}--(9.05,-4.95) (9.86,-1.93)--(9.04,-3.58) ;

\draw[thick] (9.41,-2.64)--(9.41,-2.64)node {.}  node[below left=-2pt] {\color{black}8};
\end{scope}


\begin{scope}[xshift=0cm,yshift=0cm]
\draw[green,thick](11.72,-3.86)node {.} node[left] {\color{black}9}-- (13.62,-3.33)node {.} node[below] {\color{black}10}--
(13.75,-2.35) node {.}  node[right] {\color{black}12}
--
(13.22,-1.53)
node {.}  node[above] {\color{black}\{11,13\}}
--(11.72,-3.86)--(13.75,-2.35)
(13.22,-1.53)--(13.62,-3.33)
 ;

\draw[red,thick] 
(11.72,-3.86)- -(14.10,-2.90) 
node {.}  node[right] {\color{black}14}
--(13.62,-3.33)
(14.10,-2.90) --
(13.75,-2.35) 
;

\draw[thick] (13.25,-2.28)--(13.25,-2.28)node {.}  node[below left=-2pt] {\color{black}8};
\end{scope}

\end{tikzpicture}

View from particle 2\quad View from particle 3 \quad View from particle 4\quad View from particle 5
\caption{Four projections from the viewpoint of particles 2, 3, 4 and 5, respectively, of the two bounded 4-simplices (shown in blue and red) for the topology type 4 in table \ref{458} and near the soft limit. In the strict soft limit, the two bounded chambers collapse to a point where the soft particle lies.  \label{11fourf2} }
\end{figure}

For  the topology type 4,  
 there are two bounded 4-simplices formed by the dominating 3-planes using positive kinematic data. See their projections in   figure \ref{11fourf2}.
As summarized in  table \ref{f2}, these two bounded chambers share a tetrahedron of vertices $\{9,10,11,12\}$ as a common facet, which is passed by  the dominating 3-plane determined by $\{1,2,3,6\}$. Another three dominating 3-planes 
pass  both facets of different bounded chambers.  Two dominating 3-planes  only pass a facet of either the blue or red bounded chamber.    As for the last dominating 3-plane, it just passes a dim-2 boundary determined by $\{9,10,11\}$ of the shared facet. 

 The six hard particles 1, 2, 3, 5, 6 and 7 lie on the lines determined by \{9,11\}, \{10,11\}, \{9,10\}, \{11,12\}, \{9,13\}, \{10,14\}, respectively, while particle 4  lies on the line that  passes $\{12, 13, 14\}$ at the same time.

\begin{table}[H]
\qquad\qquad\quad
\begin{tabular}{c|c|c}
${\rm Particles~ to~determine \atop dominating~3-planes}$ &${\rm Vertices ~of~ the ~facet~passed \atop by~the~ blue~chamber}$ &${\rm Vertices ~of~ the ~facet~passed \atop by~the~ red~chamber}$ \\
\hline
 \{1,2,3,6\} & \{9,10,11,12\} & \{9,10,11,12\} \\
 \{1,4,5,6\} & \{9,11,12,13\} & \{9,11,12,14\} \\
 \{2,4,5,7\} & \{10,11,12,13\} & \{10,11,12,14\} \\
 \{3,4,6,7\} & \{9,10,12,13\} & \{9,10,12,14\} \\
 \{1,2,3,5\} & \{9,10,11,13\} & - \\
 \{1,2,3,7\} & - & \{9,10,11,14\} \\
  \{1,2,3,4\} & - & - \\ 
\end{tabular}
\caption{ Dominating 3-planes and the facets they pass by in figure \ref{11fourf2}. \label{f2}} 
\end{table}

 





\bibliographystyle{JHEP}
\bibliography{references}

\end{document}